\documentclass[
,showpacs ,amssymb,nobibnotes, aps]{revtex4}
\usepackage{graphicx}
\input{epsf}

\usepackage{amsmath,amssymb}
\usepackage{bm}

\newcommand{\dalm}{\kern1pt\vbox{\hrule height 0.9pt\hbox{\vrule width 0.9pt
\hskip 2.5pt\vbox{\vskip 5.5pt}\hskip 3pt\vrule width 0.3pt}\hrule height 0.3pt}
\kern1pt}

\begin{document}



\title{Gravitational radiation from collapsing  magnetized dust}

\author{Hajime Sotani$^{1}$}\email{sotani@gravity.phys.waseda.ac.jp}
\author{Shijun Yoshida$^{2}$}\email{yoshida@heap.phys.waseda.ac.jp}
\author{Kostas D. Kokkotas$^{1}$}\email{kokkotas@astro.auth.gr}
\affiliation{
$^{1}$Department of Physics, Aristotle University of Thessaloniki, 
Thessaloniki 54124, Greece \\
$^{2}$Science and Engineering, Waseda University, Okubo, Shinjuku, 
Tokyo 169-8555, Japan}




\date{\today}

\begin{abstract}
In this article we study the influence of magnetic fields on the axial 
gravitational waves emitted during the collapse of a homogeneous dust sphere.
We found that while the energy emitted depends weakly on the initial matter perturbations it has strong dependence on the strength and the distribution of the magnetic field perturbations. 
The gravitational wave output of such a collapse can be up to an order of magnitude larger or smaller calling for detailed numerical 3D studies of collapsing magnetized configurations.

\end{abstract}

\maketitle
\section{Introduction}
\label{sec:Intro}

The direct detection of gravitational waves will be a significant breakthrough 
both for fundamental physics and astrophysics. 
With the information collected from gravitational waveforms,
one will be able to make validation of general relativity, collection of
astronomical data, and examine the nature of matter in supranuclear densities. 
Among the most important applications of the gravitational wave
observations is the asteroseismology in which seismic information for determining
stellar structures is obtained by gravitational waveforms \cite{Andersson1996,
Benhar1999,Andersson2001,Kokkotas2001,Sotani2002, Sotani2003,Sotani2004}.
Currently, several ground-based acoustic and laser
interferometric detectors for gravitational waves like LIGO, TAMA300, GEO600,
and VIRGO are in operation but there is not yet any direct detection of 
gravitational waves \cite{Barish2005}. In addition to the ground-based detectors,
there is a project to launch a Laser Interferometric Space Antenna (LISA), which is planned to be launched by the end of next decade and to operate for five years \cite{LISA}.

The nonspherical stellar collapse is one of the potential sources of
gravitational waves both for the ground-based and for space detectors. 
The ground-based interferometers target the formation of stellar
mass black holes or neutron stars
because they are most sensitive to gravitational wave frequencies in
the range 10 -- 1000 Hz. Space detectors with sensitivities ranging from 
$10^{-4}$ up to $10^{-1}$ Hz might detect signals from the creation of supermassive black-holes
 \cite{Stark1985,bs1999,ssu2001,sbss2002,ss2002,Baiotti2005}. 
There are two approaches to calculate the energy radiated away as gravitational radiation during stellar collapse; the first is via direct numerical integration of the exact Einstein and matter equations, which form a coupled nonlinear system of partial differential equations, and the second is by making use of the linear perturbation analysis. During the last decade
numerical relativity  made remarkable advances and many complicated matter and spacetime
configurations can be treated with high degree of confidence
\cite{Baumgarte2003,Duez2005,Duez2006,OAM2006,OADM2006,GR2007}.

In spite of the recent great progress in numerical relativity, 
the accurate extraction of gravitational waveforms has been proved a quite difficult task. 
The reason is that the weak gravitational waves emitted during
the stellar collapse sometimes have amplitudes similar to those of unphysical
noises due to gauge modes and/or to numerical errors. 
Linear perturbation theory is extremely efficient for such processes
since the nonradial perturbations which are responsible for the emission of 
the weak gravitational waves are separated from the nonradiative  
symmetric background.
For example, for spherically symmetric backgrounds the physical quantities 
described by the perturbations can be expanded in
terms of tensor spherical harmonics and this separation of variables simplifies 
considerably the study of the problem.
Actually, the master equations for the perturbations are reduced to simple set of coupled linear
partial differential equations, which can be evolved with extremely high accuracy.
On the other hand via this procedure one might miss certain nonlinear phenomena 
that take place at the very last stages of collapse.

Linear perturbation analysis has  been used in calculations of the
gravitational wave emission from the stellar collapse to a black hole
\cite{Cunningham1978,Seidel1987,Iguchi1998,Harada2003}. Cunningham, Price,
and Moncrief derived the perturbation equations on the Oppenheimer-Snyder
solution, which describes collapse of homogeneous dust \cite{OS1939} and
calculated gravitational radiation emitted during the collapse to a black
hole \cite{Cunningham1978}. By using the gauge invariant perturbation
formalism on the spherically symmetric spacetime formulated by Gerlach and
Sengupta \cite{Gerlach1979}, Seidel and co-workers \cite{Seidel1987}
investigated the gravitational waves from the stellar collapse in which
a neutron star is born. The gravitational waves from collapse
of an inhomogeneous dust, which can be described by nonradial perturbations
of the Lema\^{\i}tre-Tolman-Bondi solution \cite{ltb1933}, are computed by
Iguch, Nakao, and Harada \cite{Iguchi1998}. Harada, Iguchi, and Shibata
\cite{Harada2003} calculated the axial parity gravitational waves emitted
from collapse of a supermassive star to a black hole by employing the covariant
gauge-invariant formalism for nonradial perturbations on spherically
symmetric spacetime and the coordinate-independent matching conditions at
stellar surface, devised by Gundlach and Mart\'{\i}n-Garc\'{\i}a \cite{Gundlach2000}.

Although, as mentioned above, there have been many investigations of
gravitational radiation from stellar collapse with the linear perturbation
analysis, effects of magnetic fields on the gravitational radiation have not been
taken into account. While Cunningham $et$ $al$. dealt with electromagnetic
perturbations of the Oppenheimer-Snyder solution \cite{Cunningham1978},
they omitted the conduction of the fluid. In other words, Cunningham $et$ $al$.
did not consider the direct coupling between fluid and magnetic field.
However, it has been recently realized the importance of effects of magnetic fields
on the evolution of compact objects again due to the advent of high-performance
instruments like satellite-borne detectors. One of the most impressive examples
is the discovery of magnetars, which are neutron stars whose strength of magnetic
fields is estimated about $10^{15}$ G. The magnetar model, in which observed
activities are powered by decay of the strong magnetic fields of magnetars,
successfully explains activity of soft gamma-ray repeaters (SGRs). 
All the four of the known SGRs have
rotation periods of $\sim 5-8$ s, and the three of them have large period
derivatives of $\sim 10^{-10}$ s s$^{-1}$, which infer the existence of magnetic
fields of $B \approx$ (5--8) $\times 10^{14}$ G \cite{Woods2005}. 
Since there is an ultra strong magnetic field in some neutron stars, which will be
born by stellar collapse, it is natural to take into account its effect
on stellar collapse. Even if the initial magnetic field is weak, it is conceivable
that the magnetic fields of the collapsing object are, due to the magnetic flux
conservation, amplified during the collapse and will most probably affect the gravitational waves
emitted.

Another example for showing the importance of magnetic fields in the evolution of
compact objects is related to gamma-ray bursts (GRBs). The short-duration GRBs could
result from hypergiant flares of magnetars associated with the SGRs \cite{Nakar2006}
or magnetized hypermassive neutron star collapse \cite{Shibata2006}.
For long-duration GRBs, strong magnetic fields provide the excitation energy
on the required time scale and drive collimated GRB outflows in the form
of relativistic jets \cite{Meszaros1997}. Additionally, the so-called
hypermassive neutron star, which can be formed after the merger of binary neutron
stars and can be in an equilibrium state due to differential rotation, could
lead to delayed collapse by the magnetic braking and viscosity, even if
the initial magnetic field and
the viscosity are very weak \cite{Shapiro2000}. Notice that even if they are weak
initially, the magnetic fields can be amplified to the required strength by
the winding-up of weak magnetic field due to differential rotation
\cite{Spruit1999,SLSS2006}.

All the examples mentioned above indeed suggest that magnetic fields play an
important role in the stellar collapse.
In this paper, we therefore consider gravitational radiation from collapse of weakly
magnetized dust spheres to explore the effects of magnetic fields on the gravitational
waves from the stellar collapse to a black hole. For the sake of simplicity, we consider,
as the background spacetime, the Oppenheimer-Snyder solution, which describes homogeneous
dust collapse. The weak magnetic fields of the dust spheres are treated as small
perturbations around the Oppenheimer-Snyder solution. We therefore regard both
the gravitational waves and the magnetic fields as perturbations on the Oppenheimer-Snyder
solution in this study. Thus we introduced, a dimensionless quantity related to
the amplitude of gravitational perturbations, $\epsilon\sim |\delta g_{\mu\nu}|$, an another one
for the strength of the magnetic field, $\eta\sim|B/(GM^2 R^{-4})^{1/2}|$.
Here $\delta g_{\mu\nu}$ 
stands for the metric perturbation and  $B$ for the magnetic field strength while $M$ and $R$ denote
the mass and radius of the star. Note that in this work we assume that 
the Lagrangian displacement of the fluid $\xi^\mu$  satisfies the condition $|\xi^\mu|/R\sim \epsilon$.
It should be emphasized that a perturbative treatment of the magnetic fields is good enough to describe the strong magnetic fields met in magnetars because dynamics of the stellar collapse is basically governed by gravity. 
In other words, the ratio of the magnetic energy ${\cal E_M}$ to the gravitational energy ${\cal E_G}$
is sufficiently small even for magnetars, i.e. a typical value of ${\cal E_M} / {\cal E_G}$
for magnetars is approximated by ${\cal E_M} / {\cal E_G} \approx B^2/(GM^2 R^{-4})
\approx 10^{-4}\times(B/10^{16} [\text{G}])$. As for gravitational perturbations, this
paper focus on the axial parity perturbations as the first step. The axial parity
gravitational waves are treated with the covariant gauge-invariant formalism.
We further assume that the two expansion parameters, $\epsilon$ and $\eta$, satisfy
$\epsilon \sim \eta^2$. In other words, we consider the gravitational radiation directly
driven by the magnetic field of the dust sphere.

This paper is organized as follows. In \S\ref{sec:II} we make a short introduction on the orders
of the various perturbative quantities that we use, then we briefly describe the gauge-invariant 
perturbation theory, and finally we show the form of the basic equations describing the magnetic fields.
Next, in \S\ref{sec:III}, we describe the formulation for the magnetized homogeneous dust collapse 
including an analytic description of the background spacetime, the magnetic fields, 
the axial parity perturbation equations and the junction conditions at stellar surface.
In \S\ref{sec:IV}, we describe the details of 
numerical procedures employed in this study, while the code test are presented in \S\ref{sec:V}.
The results related to the efficiency of the collapsing magnetized homogeneous dust spheres in  gravitational wave emission are shown in \S\ref{sec:VI}.
The final section, \S\ref{sec:VII}, is devoted to discussion and  conclusions. In this paper, we adopt the unit of $c=G=1$, where $c$
and $G$ denote the speed of light and the gravitational constant, respectively, and
the metric signature is $(-,+,+,+)$.

\section{Perturbation Theory Primer}
\label{sec:II}

\subsection{Ordering of Perturbations}
\label{sec:II-3}

Since the energy of the magnetic field is much smaller than that of the gravity 
even for magnetars, as mentioned in the previous section, it is reasonable to 
treat the electromagnetic field in the dust sphere as small perturbations. 
The perturbed metric, $\tilde{g}_{\mu\nu}$, the perturbed four-velocity of the 
fluid, $\tilde{u}^{\mu}$, and the perturbed electromagnetic tensor, 
$\tilde{F}_{\mu\nu}$, can be expanded as:
\begin{align}
 \tilde{g}_{\mu\nu} &= g_{\mu\nu} + \delta g_{\mu\nu} + {\cal O}(\epsilon^2), \\
 \tilde{u}^{\mu}    &= u^{\mu} + \delta u^{\mu} + {\cal O}(\epsilon^2), \\
 \tilde{F}_{\mu\nu} &= F_{\mu\nu} + \delta F_{\mu\nu} + {\cal O}(\eta^2)\,,
\end{align}
where $g_{\mu\nu}$ is the background metric tensor, $u^{\mu}$, the four-velocity 
of the fluid and $F_{\mu\nu}$ the electromagnetic tensor. 
Both $g_{\mu\nu}$ and $u^{\mu}$ are defined as solutions of a collapsing 
spherical dust sphere in the absence of any electromagnetic field. 
For convenience we introduced two small dimensionless
parameters related to strength of the magnetic field and to amplitude of the 
gravitational waves, i.e., $\eta\sim|B/(GM^2 R^{-4})^{1/2}|$ and 
$\epsilon\sim |\delta g_{\mu\nu}|$. 
Moreover, we assume that the Lagrangian displacement is small i.e. $|\xi^\mu|/R\sim \epsilon$.
Finally, we need to mention that in this study, we consider an infinitely conductive fluid , i.e. we make use of  the so called ideal magnetohydrodynamic approximation.
Thus, the master equations for describing the magnetic field are given by the 
perfect conductivity condition, $\tilde{F}_{\mu\nu}\tilde{u}^{\nu}=0$, and the 
Maxwell equation 
$\tilde{F}_{\mu\nu,\alpha}+\tilde{F}_{\nu\alpha,\mu}+\tilde{F}_{\alpha\mu,\nu}=0$.
The fisrst order ($\eta^1$) form of the above two conditions will be written as: 
\begin{align}
 &\delta F_{\mu\nu}u^{\nu}=0 \,, \\
 &\delta{F}_{\mu\nu,\alpha}+\delta{F}_{\nu\alpha,\mu}+\delta{F}_{\alpha\mu,\nu}=0\,,
\end{align}
which determine the magnetic field corrections to the spherical dust sphere.
Up to this order of approximation, the variations induced by the presence of the magnetic field
do not affect the spherical symmetry of the system, since the Lorentz force 
which induces deformations in the geometry is of second order ($\eta^2$).

In a similar fashion, both the Einstein tensor and the energy-momentum tensor 
can be expanded in powers of $\epsilon$ and $\eta$ as
\begin{align}
 \tilde{G}_{\mu\nu} &= G_{\mu\nu} +  \delta G_{\mu\nu} + {\cal O}(\epsilon^2), \\
 \tilde{T}^{(M)}_{\mu\nu} &= T^{(M)}_{\mu\nu} + \delta T^{(M)}_{\mu\nu} + {\cal O}(\epsilon^2), \\
 \tilde{T}^{(EM)}_{\mu\nu} &=  \delta T^{(EM)}_{\mu\nu} + {\cal O}(\epsilon\eta^2),
\end{align}
where $T^{(M)}_{\mu\nu}$ and $T^{(EM)}_{\mu\nu}$ stand for the energy-momentum 
tensors for the fluid and for the electromagnetic fields, respectively, while 
$\delta T^{(EM)}_{\mu\nu}$ is of second order in $\eta$.
The Einstein equations of order $\eta^0\epsilon^0$ are
the evolution equations describing the unperturbed spherical dust collapse.
Here we focus in the study of the influence of magnetic field on the efficiency 
of gravitational wave emission during the collapse, and thus 
we consider only those terms of the approximation that will significant
in this study. That is we omit terms such as $\eta^0\epsilon^1$ 
and we further assume that $\epsilon \sim \eta^2$.
In this order of approximation, the Einstein equations of order $\epsilon$
are reduced to the following form
\begin{equation}
\delta G_{\mu\nu} = 8\pi\{\delta T^{(M)}_{\mu\nu}+\delta T^{(EM)}_{\mu\nu}\}
                    +{\cal O}(\epsilon^2)
                  = 8\pi\delta T_{\mu\nu} +{\cal O}(\epsilon^2) \,,
\end{equation}
which describes gravitational perturbations driven both by the magnetic field 
and the fluid motions of the collapsing dust sphere.

\subsection{Gauge-Invariant Perturbation Theory}
\label{sec:II-1}

The gauge-invariant perturbation theory for spherically symmetric background 
spacetime has been formulated by Gerlach and Sengupta \cite{Gerlach1979} 
while its covariant formulations has been developed by 
Gundlach and Mart\'{\i}n-Garc\'{\i}a \cite{Gundlach2000}. 
Here we only briefly describe this formalism for special case of axial
parity perturbations.

\subsubsection{Background Spacetime}

A spherically symmetric four dimensional spacetime ${\cal M}$ can be decomposed 
as a product of the form  ${\cal M}={\cal M}^2 \times {\cal S}^2$, 
where ${\cal M}^2$ is a 2-dimensional (1+1) reduced spacetime and 
${\cal S}^2$ a 2-dimensional spheres. 
In other words, the metric $g_{\mu\nu}$ and the stress-energy tensor $T_{\mu\nu}$
on ${\cal M}$ can be written in the form
\begin{align}
 g_{\mu\nu} &\equiv \mbox{diag} (g_{AB},R^2\gamma_{ab}), \\
 T_{\mu\nu} &\equiv \mbox{diag} (T_{AB},QR^2\gamma_{ab}),
\end{align}
where $g_{AB}$ is an arbitrary ($1+1$) Lorentzian metric on
${\cal M}^2$, $R$ a scalar on ${\cal M}^2$, $Q$ some function on ${\cal M}^2$ 
and $\gamma_{ab}$  is the unit curvature metric on ${\cal S}^2$.
Note that  if the background spacetime is spherically symmetric then $Q=T^a_{\ a}/2$ .
Here and henceforth the Greek indices denote the spacetime components, 
the capital Latin indices the ${\cal M}^2$ components, and the small Latin indices
are used to denote the ${\cal S}^2$ components. 
Furthermore, the covariant derivatives on ${\cal M}$, ${\cal M}^2$, and
${\cal S}^2$ are represented by $_{;\mu}$, $_{|A}$, and $_{:a}$, respectively.
Finally, the totally antisymmetric covariant unit tensor on ${\cal M}^2$ 
is denoted as $\varepsilon_{AB}$ and on ${\cal S}^2$  as $\varepsilon_{ab}$.

\subsubsection{Nonradial Perturbations}

As mentioned before, in this paper, we only consider axisymmetric axial parity
perturbations both for the metric $\delta g_{\mu\nu}$ and the matter
perturbations $\delta T_{\mu\nu}$, which are given by
\begin{align}
 \delta g_{\mu\nu} & \equiv   \left(
   \begin{array}{cc}
       0 &  h_A^{\rm axial} S_a^{l} \\
       h_A^{\rm axial} S_a^{l} &  h (S_{a:b}^{l} + S_{b:a}^{l})
   \end{array}\right), \label{AMP} \\
 \delta T_{\mu\nu} & \equiv   \left(
   \begin{array}{cc}
       0 &  \Delta t_A^{\rm axial} S_a^{l} \\
       t_A^{\rm axial} S_a^{l} &  \Delta t(S_{a:b}^{l} + S_{b:a}^{l})
   \end{array}\right)\,, \label{AFP}
\end{align}
where $S_{a}^{l}\equiv\varepsilon^{b}_{\ a}P_{l:b}$ while $P_l$ stands for the 
Legendre polynomial.
The gauge-invariant variables of the perturbations are then defined as
\begin{align}
 k_A &\equiv h_A^{\rm axial}-h_{|A}+2hv_A, \\
 L_A &\equiv \Delta t_A^{\rm axial} - Qh_A^{\rm axial}, \\
 L   &\equiv \Delta t   - Qh,
\end{align}
where $v_A\equiv R_{|A}/R$ \cite{Gundlach2000}.
In terms of the gauge-invariant variables, the master equations for the axial 
parity perturbations
are given by
\begin{gather}
  k_A^{\ |A} = 16\pi L, \label{BE1} \\
  -\left[R^4\left(\frac{k^A}{R^2}\right)^{|C} - 
         R^4\left(\frac{k^{C}}{R^2}\right)^{|A}\right]_{|C}
    + (l-1)(l+2)k^A = 16\pi R^2 L^A, \label{BE2}
\end{gather}
\begin{equation}
 (R^2 L^A)_{|A} = (l-1)(l+2)L. \label{BE3}
\end{equation}

\subsection{Basic equations for the magnetic field}
\label{sec:II-2}

As mentioned earlier, the electromagnetic field perturbations, $\delta F_{\mu\nu}$, 
are governed by the Maxwell equations, i.e.
\begin{gather}
\delta F_{\mu\nu,\sigma} +\delta F_{\nu\sigma,\mu} +\delta F_{\sigma\mu,\nu} = 0, 
\label{Maxwell-1} \\
\delta F^{\mu\nu}_{\ \ ;\nu} = 4\pi \delta J^{\mu}, \label{Maxwell-2}
\end{gather}
where $\delta J^{\mu}$ is the perturbations of the current four-vector. 
Note that equations (\ref{Maxwell-1}) and (\ref{Maxwell-2}) are correct up to 
order of $\eta^1\epsilon^0$.
The perturbation of the electromagnetic field energy-momentum tensor, 
$\delta T_{\mu\nu}^{(EM)}$,  in this order of approximation has the form 
\begin{equation}
\delta T_{\mu\nu}^{(EM)} = \frac{1}{4\pi} \left(\delta F_{\mu\alpha}
\delta F_{\nu\beta}g^{\alpha\beta}
     - \frac{1}{4}g_{\mu\nu}\delta F_{\alpha\beta}
     \delta F_{\lambda\gamma}g^{\alpha\lambda}
g^{\beta\gamma}\right).
\end{equation}
The electric  $E_{\mu}$ and the magnetic field $B_{\mu}$ associated 
with the four-velocity of the fluid $u^{\nu}$ are defined as
\begin{align}
 E_{\mu} &= \delta F_{\mu\nu}u^{\nu}, \\
 B_{\mu} &= \frac{1}{2}\varepsilon_{\mu\nu\alpha\beta}u^{\nu}\delta F^{\alpha\beta}\,.
\end{align}
Finally, we remind that in this paper we consider infinitely conductive fluids,
i.e. the ideal magnetohydrodynamic approximation has been adopted, 
according to which $E_{\mu}=\delta F_{\mu\nu}u^{\nu}=0$, where
$u^{\nu}$ is the unperturbed four-velocity of the infinitely conductive fluid.

\section{Magnetized Homogeneous Dust Collapse: Formulation}
\label{sec:III}

\subsection{Background spacetime for perturbations}
\label{sec:III-1}

Here we briefly describe the background spacetime which will be later 
endowed with a  magnetic field.
We consider perturbations around a homogeneous spherically symmetric dust 
collapse described by the Oppenheimer-Snyder (OS) solution, whose line 
element inside the dust sphere is given by
\begin{align}
ds^2 
  &=g_{\mu\nu}dx^\mu dx^\nu \,, \nonumber \\
  &=-d\tau^2+R^2(\tau)[d\chi^2 + \sin^2 \chi (d \theta^2 +\sin^2\theta d \phi^2)]\,,
\label{OSM} \\
  &= R^2(\eta)[-d\eta^2+d\chi^2+ \sin^2 \chi (d \theta^2 +\sin^2\theta d \phi^2)]\,,
\label{OSM1}
\end{align}
where $\chi$ is a radial coordinate defined in the range of 
$0\le\chi\le\chi_0$. Here $\chi_0$ is the stellar surface and it is
assumed that $\chi_0<\pi/2$.
In the line element defined earlier, $R(\eta)$ is the scale factor 
and $\tau(\eta)$ is the proper time of an observer comoving with the fluid, 
defined in terms of the conformal time $\eta$ as follows
\begin{align}
 R(\eta)    &= \frac{M}{\sin^3 \chi_0}(1 + \cos\eta)\,, \\
 \tau(\eta) &= \frac{M}{\sin^3 \chi_0}(\eta + \sin\eta)\,,
\end{align}
where $M$ is the total gravitational mass of the dust sphere.
The energy-momentum tensor for the dust fluid is written as
\begin{equation}
 T_{\mu\nu}^{(M)} = \rho u_{\mu}u_{\nu},
\end{equation}
where $\rho$ is the rest mass density given by
\begin{equation}
 \rho (\eta) = \frac{3 \sin^6\chi_0}{4\pi M^2}(1+\cos\eta)^{-3}\,,
\end{equation}
and $u^{\mu}$ denotes the four-velocity of the dust, descibed in  terms of
comoving coordinates as
%
\begin{equation}
 u^\mu=\delta^\mu_{\ \tau}\, \ \ \ \mbox{or}\ \ \ u^\mu=R(\eta)\delta^\mu_{\ \eta}
\end{equation}
where $\delta^\mu_{\ \nu}$ means the Kronecker delta.
The spacetime outside the dust sphere is described by the Schwarzschild metric, 
i.e.,
\begin{equation}
 ds^2 = -f(r)dt^2 + f(r)^{-1}dr^2 + r^2(d \theta^2 + \sin^2\theta d \phi^2), \label{Sch}
\end{equation}
where $f(r)\equiv 1-2M/r$. 
From the junction conditions at the surface of the dust sphere, we obtain
the relationships between the $(\eta,\chi)$-coordinates and the $(t,r)$-coordinates, given by
\begin{align}
 r_s &= R(\eta)\sin\chi_0, \\
 \frac{t}{2M} &= \ln \left|\frac{[(r_{s0}/2M)-1]^{1/2}+\tan(\eta/2)}{[(r_{s0}/2M)-1]^{1/2}-\tan(\eta/2)}\right|
    + \left(\frac{r_{s0}}{2M}-1\right)^{1/2}\left[\eta + \left(\frac{r_{s0}}{4M}\right)(\eta + \sin\eta)\right],
\end{align}
where $r_{s0}\equiv r_s(t=0) = 2M/\sin^2\chi_0$ is the initial stellar radius 
in Schwarzschild coordinates.

\subsection{The Magnetic field of the star}
\label{sec:III-2}

As mentioned earlier, we consider weakly magnetized dust spheres in which the 
magnetic effects on the dust fluid are treated as small perturbations on the 
OS solution. Moreover, for the sake of simplicity we assume that
the electromagnetic fields are axisymmetric. Thus, perturbations of the
electromagnetic fields, $\delta F_{\mu\nu}$, and the current four-vector, 
$\delta J_{\mu}$, can be described in terms of the Legendre polynomial 
$P_{l_M}$ by the following relations
\begin{align}
 \delta F_{03} &=  - \delta F_{30} = e_1 \sin\theta \partial_{\theta} P_{l_M}\,, \label{F-1} \\
 \delta F_{13} &=  - \delta F_{31} = b_1 \sin\theta \partial_{\theta} P_{l_M}\,, \label{F-2} \\
 \delta F_{23} &=  - \delta F_{32} = b_2 \sin\theta P_{l_M}\,, \label{F-3} \\
 \delta F_{01} &=  - \delta F_{10} = e_2 P_{l_M}\,, \label{F-4} \\
 \delta F_{02} &=  - \delta F_{20} = e_3 \partial_{\theta} P_{l_M}\,,\label{F-5} \\
 \delta F_{12} &=  - \delta F_{21} = b_3 \partial_{\theta} P_{l_M}\,, \label{F-6} \\
%
\delta J_{\mu} &= \left(j_A P_{l_M},j^{(p)}P_{l_M:a}+j^{(a)}S^{l_M}_a \right)\, .
\end{align}
Notice, that here we have used $l_M$ to denote the angular quantum number with respect to the electromagnetic fields
to discriminate it from the one for the gravitational waves $l$.

In the interior of the dust sphere, the perfect conductivity condition 
$\delta F_{\mu\nu}u^\nu=0$ is reduced into $\delta F_{0\mu}=0$.  
This assumption leads to the following simplifications
\begin{equation}
e_1 = e_2 = e_3 =0\,.
\label{Maxwell1-0in}
\end{equation}
By direct substitution  of equations (\ref{F-1}) through (\ref{F-6}) into the
Maxwell equation (\ref{Maxwell-1}), we obtain the basic equations  
describing the magnetic fields, which have the following simple form 
\begin{gather}
 \partial_{\eta} b_1 = \partial_{\eta} b_2 = \partial_{\eta} b_3 = 0, \label{Maxwell1-1in} \\
 l_M(l_M+1)b_1 + \partial_{\chi} {b_2} = 0\,. \label{Maxwell1-2in}
\end{gather}
The first of theses relations, equation (\ref{Maxwell1-1in}), suggests that 
a comoving observer does not observe any change in the magnetic field 
distributions. 
All the components of the electromagnetic fields can be determined through 
equations (\ref{Maxwell1-0in}), (\ref{Maxwell1-1in}), and (\ref{Maxwell1-2in}).
The Maxwell equation (\ref{Maxwell-2}) that we still have not use can be 
regarded as the definition of the current four-velocity. 
This implies that the perturbations of the current four-velocity can be written 
as follows:
\begin{align}
 j_{\eta} &= 0, \label{Maxwell2-1in} \\
 j_{\chi} &= -\frac{l_M(l_M+1)}{4 \pi}\frac{b_3}{R^2 \sin^2\chi}, \label{Maxwell2-2in} \\
 j^{(p)}  &= -\frac{\partial_{\chi}b_3}{4\pi R^2}, \label{Maxwell2-3in} \\
 j^{(a)}  &= -\frac{1}{4\pi R^2}\left(\partial_{\chi}b_1 + \frac{b_2}{\sin^2\chi}\right)\,.
\label{Maxwell2-4in}
\end{align}
The electromagnetic fields outside the star are also given by similar expressions 
to equations (\ref{F-1})--(\ref{F-6}). 
In order to avoid mixing of the various quantities inside and  outside the star 
we will indicate the ones in the exterior with a tilde i.e.
$\tilde{e}_1$, $\tilde{b}_1$, and so on.
Since the exterior of the star is vacuum, we cannot make use
of the perfect conductivity condition there. 
Instead, we make an alternative assumption that is we demand the vanishing of the 
current perturbations outside the star i.e. $\delta J^{\mu}=0$.  
This assumption simplifies considerably the Maxwell equation (\ref{Maxwell-2}) 
leading into thre following set of equations
\begin{gather}
 \partial_{r_*}(r^2\tilde{e}_2) - l_M(l_M+1)\tilde{e}_3 = 0, \label{Maxwell2-1out} \\
 \partial_t(r^2\tilde{e}_2) - l_M(l_M+1)f\tilde{b}_3 = 0, \label{Maxwell2-2out} \\
 \partial_t \tilde{e}_1 - \partial_{r_*}(f\tilde{b}_1) - \frac{f}{r^2}\tilde{b}_2 = 0, \label{Maxwell2-3out}
\end{gather}
where $r_*$ is the tortoise coordinate defined as $r_* = r + 2M\ln(r/2M-1)$. 
Furthermore, from equation (\ref{Maxwell-1}), we get a set of equations similar  
to equations (\ref{Maxwell1-1in}) and (\ref{Maxwell1-2in}),
given by
\begin{gather}
 \tilde{e}_2 - \partial_r \tilde{e}_3 + \partial_t \tilde{b}_3 = 0, \label{Maxwell1-1out} \\
 l_M(l_M+1)\tilde{e}_1 + \partial_t \tilde{b}_2 = 0, \label{Maxwell1-2out} \\
 l_M(l_M+1)\tilde{b}_1 + \partial_r \tilde{b}_2 = 0. \label{Maxwell1-3out}
\end{gather}
The six equations for the electromagnetic fields, 
i.e. equations (\ref{Maxwell2-1out}) through (\ref{Maxwell1-3out}), can be
reduced to two decoupled wave equations
\begin{gather}
 -\partial_t^2\tilde{b}_2 + \partial_{r_*}^2 \tilde{b}_2 - \frac{l_M(l_M+1)}{r^2}f\tilde{b}_2 = 0\,, \label{EM-axial} \\
 -\partial_t^2(r^2\tilde{e}_2) + \partial_{r_*}^2 (r^2\tilde{e}_2) - \frac{l_M(l_M+1)}{r^2}f(r^2\tilde{e}_2) = 0\,.
    \label{EM-polar}
\end{gather}
Moreover, these two wave equations can be rewritten in terms of the double null coordinates,
$\tilde{u}=t-r_*$ and $\tilde{v}=t+r_*$, as
\begin{gather}
 \frac{\partial^2 \tilde{b}_2}{\partial \tilde{u} \partial \tilde{v}} + \frac{l_M(l_M+1)}{4r^2}f\tilde{b}_2 = 0,
     \label{exterior_EM1} \\
 \frac{\partial^2 (r^2 \tilde{e}_2)}{\partial \tilde{u} \partial \tilde{v}}
     + \frac{l_M(l_M+1)}{4r^2}f (r^2 \tilde{e}_2) = 0.
     \label{exterior_EM2}
\end{gather}
At the surface of the star, we implement the following junction conditions 
for the electromagnetic field
\begin{align}
 n^{\mu}B_{\mu} &= \tilde{n}^{\mu}\tilde{B}_{\mu}, \\
 q_{\mu}^{\ \nu}E_{\nu} &= \tilde{q}_{\mu}^{\ \nu}\tilde{E}_{\nu},
\end{align}
where $n^{\mu}$, $\tilde{n}^{\mu}$ are the unit outward normal vector to the 
stellar surface defined in the interior and the exterior coordinates, 
respectively, while $q_{\mu}^{\ \nu}$, $\tilde{q}_{\mu}^{\ \nu}$ are the 
corresponding projection tensors associated with $n^{\mu}$ and $\tilde{n}^{\mu}$.
Therefore the junction conditions reduced to the following set of relations
\begin{align}
 b_2=\tilde{b}_2, \hspace{1cm}
 \tilde{e}_1 + \frac{\tilde{u}^1}{\tilde{u}^0}\tilde{b}_1 = 0, \hspace{1cm}
 \tilde{e}_3 + \frac{\tilde{u}^1}{\tilde{u}^0}\tilde{b}_3 = 0,
\label{magcon}
\end{align}
where
\begin{equation}
 \frac{\tilde{u}^1}{\tilde{u}^0} = \frac{\partial_{\eta}R}{R}f\tan\chi_0.
\end{equation}

Concluding we mention that in this paper we focus only on dipole 
electromagnetic fields, i.e., electromagnetic fields associated
with $l_M = 1$. 
Observations actually are in favor of the existence of dipole electromagnetic 
fields and moreover these fields can  drive the quadrupole gravitational 
radiation as we will see in the next section.
Finally, the formalism developed here accounts for electromagnetic fields which
lie both inside and outside the star, in this study as a first step, 
we take into account only magnetic fields confined in the stellar interior. 
We therefore assume that $\tilde{e}_i = \tilde{b}_i = 0$ for $i=1,2,3$.

\subsection{Basic equations for the axial parity perturbations}
\label{sec:III-3}

\subsubsection{Interior region of the star}

As we mentioned earlier we will use the gauge-invariant formulation in the treatment
of perturbations of the OS spacetime.
The gauge-invariant form of axial perturbation equations (\ref{BE1}) and (\ref{BE2}) 
for the OS spacetime is reduced to the following set of equations
\begin{gather}
 -\partial_{\eta} k_{\eta} + \partial_{\chi} k_{\chi} - 16\pi R^2L = 0, \label{BEE1} \\
 \partial_{\chi}(R^2 \Pi\sin^4\chi) + (l-1)(l+2) \frac{k_{\eta}}{R^2} - 16\pi L_{\eta} \sin^2\chi = 0, \label{BEE2} \\
 \frac{1}{R^2}\partial_{\eta} (R^4 \Pi\sin^4\chi) + (l-1)(l+2)\frac{k_{\chi}}{R^2}
      - 16\pi L_{\chi} \sin^2\chi = 0, \label{BEE3}
\end{gather}
where $\Pi$ is the gauge-invariant quantity, defined as
\begin{equation}
  \Pi = \frac{1}{R^2} \left[\partial_{\eta}\left(\frac{k_{\chi}}{R^2\sin^2\chi}\right)
       - \partial_{\chi}\left(\frac{k_{\eta}}{R^2\sin^2\chi}\right)\right].
\end{equation}
The regularity condition at the stellar center, suggests the introduction 
of a new function $\bar{\Pi}$ defined as
\begin{equation}
 \Pi=(R\sin\chi)^{l-2}\bar{\Pi}\,, \label{eq:regularity}
\end{equation}
which is analytic at the stellar center.
By using equations (\ref{BEE2}) and (\ref{BEE3}), one can derive a single wave equation for
$\bar{\Pi}$, given by
\begin{align}
 -\partial_{\eta}^2 \bar{\Pi} + \partial_{\chi}^2 \bar{\Pi}
     + 2(l+1)\left(\frac{\cos\chi}{\sin\chi}\partial_{\chi}\bar{\Pi}
     - \frac{\partial_{\eta}R}{R}\partial_{\eta}\bar{\Pi}\right)
     - \frac{(2l-1)(l+2)R(0)}{2R}\bar{\Pi}
     = \frac{16\pi}{R^l\sin^l\chi}(\partial_{\chi}L_{\eta} - \partial_{\eta}L_{\chi}), \label{WE}
\end{align}
where $R(0) = 2M/\sin^3\chi_0$.
Moreover the equation of motion (\ref{BE3}) is rewritten as
\begin{equation}
  -\partial_{\eta}(R^2 L_{\eta}\sin^2\chi) + \partial_{\chi} (R^2 L_{\chi} \sin^2\chi)
     = (l-1)(l+2)R^2 L. \label{BEE4}
\end{equation}

The perturbation of the energy-momentum tensor $\delta T_{\mu\nu}$ can be splitted 
into two parts as shown before:
\begin{equation}
 \delta T_{\mu\nu} = \delta T_{\mu\nu}^{(M)} + \delta T_{\mu\nu}^{(EM)},
\end{equation}
where $ T_{\mu\nu}^{(M)}$ and $T_{\mu\nu}^{(EM)}$ are the energy-momentum tensors
for the dust  and the electromagnetic field, respectively. 
Since axial parity perturbations of the four-velocity of the fluid, 
$\delta u_{\mu}$, defined as
\begin{equation}
\delta u_{\mu} = (0,0,\beta(\tau,\chi)S_{a}^{l})\,,
\end{equation}
the expansion coefficients of $\delta T_{\mu\nu}^{(M)}$ introduced in equation 
(\ref{AFP}) are given by
\begin{gather}
 \Delta t_{\eta}^{(M)} = \beta\rho u_{\eta}, \\
 \Delta t_{\chi}^{(M)} = \Delta t^{(M)} = 0.
\end{gather}

In this paper we constrain our study to the quadrupole gravitational radiation 
emitted by axial parity perturbations. 
The reason is that quadrupole radiation directly couples with dipole magnetic 
fields and it is dominant component for gravitational wave emission. 
As discussed before, the perturbations of the
energy-momentum tensor for the electromagnetic field, $\delta T_{\mu\nu}^{(EM)}$, 
are of order  $\sim \left(\delta F_{\mu\nu}\right)^2$. 
Therefore, we cannot  achieve separation of variables for
$\delta T_{\mu\nu}^{(EM)}$ even if the background spacetime is spherically symmetric. 
Since we assume dipole electromagnetic fields ($l_{M}=1$), then 
$\delta T_{\mu\nu}^{(EM)}$ will contain terms associated with $l=0$ and $l=2$. 
For the detailed calculations
of various components of $\delta T_{\mu\nu}^{(EM)}$ inside the star, see Appendix \ref{sec:appendix_1}. 
Finally, the expansion coefficients for
$\delta T_{\mu\nu}^{(EM)}$ associated with $l=2$ are given by
\begin{align}
 \Delta t_{\eta}^{(EM)} &= 0, \\
 \Delta t_{\chi}^{(EM)} &= -\frac{b_2 b_3}{12\pi R^2\sin^2\chi}, \\
 \Delta t^{(EM)}        &= \frac{b_1 b_3}{12\pi R^2}.
\end{align}
Thus we can derive the gauge-invariant quantities for the total matter perturbations (the dust fluid and the magnetic field), $L_A$ and $L$, which have the following form
\begin{align}
 L_{\eta} &= -R\beta\rho\,, \label{L1} \\
 L_{\chi} &= -\frac{b_2 b_3}{12\pi R^2 \sin^2\chi}\,, \label{L2}\\
 L        &= \frac{b_1 b_3}{12\pi R^2}\,. \label{L3}
\end{align}
Substituting equations (\ref{L1}) through (\ref{L3}) into equation (\ref{BEE4}), we get
the following equation of motion for $\beta$ 
\begin{align}
 \partial_{\eta} \beta = \frac{1}{12\pi R^3\rho \sin^2\chi}
     \left[b_2 (\partial_{\chi}b_3) + \left\{1 - \frac{(l-1)(l+2)}{2}\right\}b_3 (\partial_{\chi}b_2)\right],
     \label{beta_eta}
\end{align}
notice that in this derivation we have used the perturbed Maxwell equation (\ref{Maxwell1-2in}). 
Using the relations, $R^3\rho = 3R(0)/8\pi = const.$, $\partial_{\eta}b_2=0$, 
and $\partial_{\eta}b_3=0$, we can analytically integrate equation 
(\ref{beta_eta}) with respect to conformal time $\eta$, to get the solution
\begin{equation}
 \bar{\beta}(\eta,\chi) = \frac{2\eta}{9R(0)R^{l+1}\sin^{l+3}\chi}
 \left[b_2 (\partial_{\chi}b_3)
     + \left\{1 - \frac{(l-1)(l+2)}{2}\right\}b_3 (\partial_{\chi}b_2)\right]
     + \left(\frac{R(0)}{R}\right)^{l+1}\bar{\beta}_0(\chi), \label{interior_GW3}
\end{equation}
where $\bar{\beta}=\beta/(R\sin\chi)^{l+1}$ and $\bar{\beta}_0(\chi)$ is the 
initial distribution of $\bar{\beta}$. 
Following Ref. \cite{Harada2003}, in this paper, we adopt three different 
definitions for the initial distribution $\bar{\beta}_0(\chi)$ i.e.
\begin{align}
 \bar{\beta}_0(\chi) &= {\cal U}_1 (= const.), \label{beta01} \\
 \bar{\beta}_0(\chi) &= {\cal U}_2 \exp\left[-\left(\frac{R(0)\sin\chi}{R_c}\right)^2\right], \label{beta02} \\
 \bar{\beta}_0(\chi) &= {\cal U}_3 \exp\left[-\left(\frac{R(0)\sin\chi - r_{s0}}{R_c}\right)^2\right], \label{beta03}
\end{align}
where ${\cal U}_i$ is an arbitrary constant and $R_c$ is a scale factor describing 
the inhomogeneity of the fluid velocity's initial deistribution. 
Here following Ref. \cite{Harada2003} we choose $R_c=r_{s0}/3$.

In the actual numerical calculations, we use the two null coordinates, 
($u = \eta - \chi$ and $v = \eta + \chi$) and the master equation (\ref{WE}) is 
rewritten  in these coordinates as
\begin{gather}
 \frac{\partial^2 \bar{\Pi}}{\partial u\partial v}
     + \frac{l+1}{2}\left(\frac{\cos\chi}{\sin\chi} + \frac{\partial_{\eta}R}{R}\right)\frac{\partial \bar{\Pi}}{\partial u}
     - \frac{l+1}{2}\left(\frac{\cos\chi}{\sin\chi} - \frac{\partial_{\eta}R}{R}\right)\frac{\partial \bar{\Pi}}{\partial v}
     + \frac{(2l-1)(l+2)R(0)}{8R}\bar{\Pi} = S(\eta,\chi),
     \label{interior_GW1} \\
  S(\eta,\chi) = 4\pi R^2 \rho \left\{(l+1)\cos\chi \bar{\beta} + \sin\chi (\partial_{\chi}\bar{\beta})\right\}
     + \frac{2b_2 b_3 (\partial_{\eta}R)}{3R^{l+3}\sin^{l+2}\chi}.
     \label{interior_GW2}
\end{gather}
In summary, inside the star, our basic equations describing the gravitational perturbations are equations (\ref{interior_GW1}) and (\ref{interior_GW2}). The source term $S(\eta,\chi)$ is given uniquely by the function $\bar{\beta}(\eta,\chi)$, shown in equation (\ref{interior_GW3}), 
for a given initial distribution of the electromagnetic field and the fluid velocity perturbations.

%
\subsubsection{Exterior region of the star}
%

The Oppenheimer-Snyder solution (\ref{OSM1}) for the interior is matched with the Schwarzschild solution (\ref{Sch}) for the exterior and the master equations for perturbations 
(\ref{BE1}) and (\ref{BE2}) in the interior reduce to the well known Regge-Wheeler equation in the exterior, which has the form
\begin{gather}
 -\partial_t^2 \tilde{\Phi} + \partial_{r_*}^2 \tilde{\Phi} - \tilde{V}(r) \tilde{\Phi}
    = 16\pi r (\partial_{r_*}\tilde{L}_t - f\partial_t \tilde{L}_r), \label{WE-out} \\
 \tilde{V}(r) = f\left(\frac{l(l+1)}{r^2}-\frac{6M}{r^3}\right),
\end{gather}
where $\tilde{\Phi}$ is the Regge-Wheeler function, related to the gauge-invariant
variable $\tilde{\Pi}$ through the relationship
\begin{equation}
\tilde{\Phi}=r^3 \tilde{\Pi}=
r^3 \left[\partial_t\left(\frac{\tilde{k}_r}{r^2}\right)
    - \partial_r\left(\frac{\tilde{k}_t}{r^2}\right)\right].
\end{equation}

%
If the electromagnetic fields do not vanish outside the star, we need take into 
account their influence on the gravitational radiation emmitted during the collapse.
In the exterior the only non-vanishing contribution to the perturbations of the 
energy-momentum tensor $\delta \tilde{T}_{\mu\nu}$ is the one from the
electromagnetic field. 
The expansion coefficients for $\delta \tilde{T}_{\mu\nu}$ associated with 
the $l=2$ axial parity perturbations are then given by the following formulae
\begin{align}
 \tilde{L}_{t} &= \Delta \tilde{t}_{t}^{(EM)} = -\frac{1}{12\pi} \left(f \tilde{e}_2 \tilde{b}_1
    + \frac{1}{r^2} \tilde{e}_3 \tilde{b}_2\right), \\
 \tilde{L}_{r} &= \Delta \tilde{t}_{r}^{(EM)} = -\frac{1}{12\pi} \left(\frac{1}{f} \tilde{e}_1 \tilde{e}_2
    + \frac{1}{r^2} \tilde{b}_2 \tilde{b}_3\right), \\
 \tilde{L} &= \Delta \tilde{t}^{(EM)} = -\frac{1}{12\pi} \left(\frac{1}{f} \tilde{e}_1 \tilde{e}_3
    - f \tilde{b}_1 \tilde{b}_3\right).
\end{align}
Actually, in Appendix \ref{sec:appendix_2}, we present the detailed form of the various
components of the perturbed energy-momentum tensor for the electromagnetic field
outside the star. 

By using the vacuum Maxwell equations (\ref{Maxwell2-1out})--(\ref{Maxwell1-3out}), 
we can rewrite $\tilde{L}_{t}$, $\tilde{L}_{r}$, and $\tilde{L}$ in terms of 
$\tilde{e}_2$ and $\tilde{b}_2$ as
\begin{gather}
 \tilde{L}_{t} = \frac{f}{24\pi}\left[\tilde{e}_2 (\partial_r \tilde{b}_2)
    - \frac{2}{r} \tilde{e}_2 \tilde{b}_2 - \tilde{b}_2 (\partial_{r} \tilde{e}_2)\right], \\
 \tilde{L}_{r} = \frac{1}{24\pi f} \left[\tilde{e}_2 (\partial_t \tilde{b}_2)
    - \tilde{b}_2 (\partial_t \tilde{e}_2)\right], \\
 \tilde{L}        = \frac{1}{48\pi}\left[2r \tilde{e}_2 (\partial_t \tilde{b}_2)
    + r^2 (\partial_t \tilde{b}_2)(\partial_r \tilde{e}_2)
    - r^2 (\partial_t \tilde{e}_2)(\partial_r \tilde{b}_2)\right].
\end{gather}
Finally, the Regge-Wheeler equation (\ref{WE-out}) can be
rewritten in terms of the double null coordinates, $\tilde{u}=t-r_*$ 
and $\tilde{v}=t+r_*$, as
\begin{equation}
 \frac{\partial^2 \tilde{\Phi}}{\partial \tilde{u} \partial \tilde{v}} + 
      \frac{1}{4}\tilde{V}(r) \tilde{\Phi}
    = \frac{f}{3r^2}\left[r^2 \tilde{e}_2 (\partial_{r_*} \tilde{b}_2)
    - \tilde{b}_2 (\partial_{r_*} r^2 \tilde{e}_2)\right]\,,
    \label{exterior_GW}
\end{equation}
where we have modified the right-hand side of the Regge-Wheeler equation by using equations (\ref{EM-axial}) and (\ref{EM-polar}). 
Still, since in this study  we don't take into account the  influence of the electromagnetic field outside the star,  the right-hand side of equation (\ref{exterior_GW}) vanishes.

\subsection{Junction conditions at the stellar surface}
\label{sec:III-4}

In order to ensure that the spacetime is regular at the stellar surface ($\chi=\chi_0$), 
we impose three junction conditions for the case of axial parity perturbations; 
first we demand the continuity of $\Pi$,
second that 
$n^A \Pi_{|A} - 16\pi(R\sin\chi)^{-2}u^A L_A = \tilde{n}^A \tilde{\Pi}_{|A} - 16\pi r^{-2} \tilde{u}^A \tilde{L}_A$,
and the last is $u^A \Pi_{|A} = \tilde{u}^{A} \tilde{\Pi}_{|A}$.
These boundary conditions arise from the continuity conditions for the 
induced metric and the exrinsic curvature \cite{Gundlach2000}.
Therefore the junction conditions are explicitly given by
\begin{gather}
 \Pi = \tilde{\Pi}, \\
 -Z + W + \frac{16\pi\beta\rho}{R\sin^2\chi_0}
     = \left((\partial_{\eta}R)\sin\chi_0 - R\cos\chi_0\right) \frac{\tilde{Z}}{f}
     + \left((\partial_{\eta}R)\sin\chi_0 + R\cos\chi_0\right) \frac{\tilde{W}}{f} \nonumber \\
     - \frac{16\pi}{fr^2}\left(\tilde{L}_t R\cos\chi_0 + f\tilde{L}_r(\partial_{\eta}R)\sin\chi_0\right),
     \label{junction1} \\
 Z + W = \left(R\cos\chi_0 - (\partial_{\eta}R)\sin\chi_0\right)\frac{\tilde{Z}}{f}
     + \left(R\cos\chi_0 + (\partial_{\eta}R)\sin\chi_0\right)\frac{\tilde{W}}{f},
     \label{junction2}
\end{gather}
where $Z = \partial \Pi / \partial u$, $W = \partial \Pi / \partial v$,
$\tilde{Z} = \partial \tilde{\Pi} / \partial \tilde{u}$, and
$\tilde{W} = \partial \tilde{\Pi} / \partial \tilde{v}$.
The last  two conditions, (\ref{junction1}) and (\ref{junction2}), can be rewritten as
\begin{align}
 W &= \left(R\cos\chi_0 + (\partial_{\eta}R)\sin\chi_0\right)\frac{\tilde{W}}{f}
     - \frac{8\pi}{fr^2}\left(\tilde{L}_t R\cos\chi_0 + f\tilde{L}_r(\partial_{\eta}R)\sin\chi_0\right)
     - \frac{8\pi\beta\rho}{R\sin^2\chi_0}, \\
 Z &= \left(R\cos\chi_0 - (\partial_{\eta}R)\sin\chi_0\right)\frac{\tilde{Z}}{f}
     + \frac{8\pi}{fr^2}\left(\tilde{L}_t R\cos\chi_0 + f\tilde{L}_r(\partial_{\eta}R)\sin\chi_0\right)
     + \frac{8\pi\beta\rho}{R\sin^2\chi_0}.
\end{align}

\section{Numerical procedure}
\label{sec:IV}

\begin{center}
\begin{figure}[htbp]
\includegraphics[height=7cm]{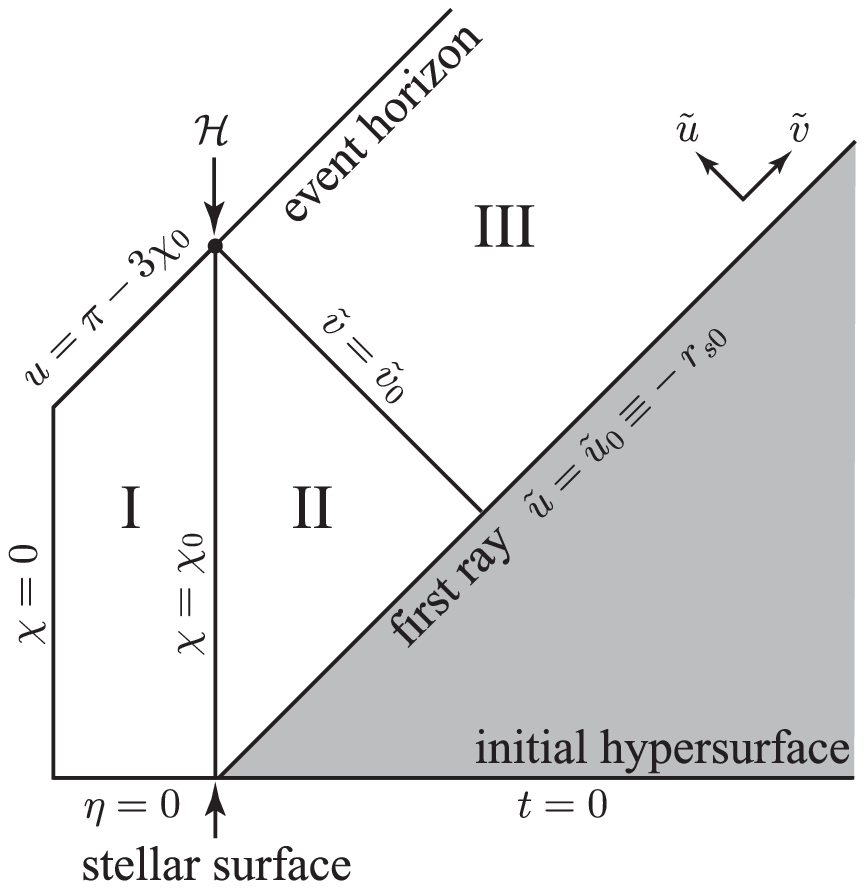}
\caption{
A schematic description of the Oppenheimer-Snyder spacetime for the collapsing 
model in characteristic coordinates.
Region I denotes the stellar interior while regions II and III correspond to
the exterior. The stellar surface, where $r=r_s$ or $\chi=\chi_0$, is the boundary between regions I and II, and the stationary region outside star is indicated by gray.
}
\label{fig-calculation-region}
\end{figure}
\end{center}

In this section we describe the numerical procedures that we will follow and the way that we generate initial data.
In order to simplify the numerical procedure and to set the initial data both in the interior and exterior of the collapsing configuration, we divide the background spacetime into three regions
named I, II, and III, as illustrated in Fig. \ref{fig-calculation-region}. 
Region I represents  the stellar interior, while regions II and III the exterior spacetime. 
Region III is separated from region II via the null hypersurface defined 
by $\tilde{v}=\tilde{v}_0$, which is generated by the ingoing null rays
emitted from the point where the stellar surface reaches the event horizon. 
Note that it is sufficient to consider the regions I, II, and III because the 
gray area in Fig. \ref{fig-calculation-region} is causally disconnected from 
the stellar interior at $\eta=0$ when the magnetic fields confined inside the star.
To solve the wave equation numerically, we make use of the finite differencing 
scheme proposed by Hamad\'{e} and Stewart \cite{Hamade1996}, in which the double 
null coordinates $(u,v)$  are employed in region I and  $(\tilde{u},\tilde{v})$
in regions II and III, respectively. 
Notice that we integrate the wave equation in region I by using a first order 
finite differencing scheme to avoid numerical instabilities appearing near the 
stellar center, while Hamad\'{e} and Stewart's original scheme is of a second 
order finite differencing scheme.

%

\subsection{Initial Data}
\label{sec:IV-1}

In order to initiate the numerical calculations, we need to provide a data 
set on the initial hypersurface for the quantities 
$\bar{\Pi}$, $\partial_u \bar{\Pi}$, $\partial_v \bar{\Pi}$, $\bar{\beta}_0$,
$b_2$, and $b_3$ for the stellar interior, and 
$\tilde{\Phi}$, $\partial_{\tilde{u}}\tilde{\Phi}$,
and $\partial_{\tilde{v}} \tilde{\Phi}$ for the stellar exterior. 
Following \cite{Cunningham1978}, we assume that the initial perturbations are 
``momentarily static''. 
Outside the star, the momentarily static initial condition for the metric 
perturbations is given as a static vacuum solution of
equation (\ref{WE-out}) in terms of hypergeometric functions
\begin{equation}
 \tilde{\Phi}_{\rm static} = 
 \frac{q_l}{l(l+1)}\left(\frac{2M}{r}\right)^{l} F_l \left(l-1,l+3,2l+2;\frac{2M}{r}\right),
     \label{static-solution}
\end{equation}
where 
$q_l$ is a constant representing the multipole moment of the star. 
Here we assume that $q_l=2M$.
We remind that there is no electromagnetic field outside the star.
Since it is a static solution, the initial perturbation outside 
the star (\ref{static-solution}) does not evolve until a light signal from 
the stellar interior arrives there.
Finally, set the initial data i.e. the static solution (\ref{static-solution}), 
on the null hypersurface $\tilde{u}=\tilde{u}_0$ for the characteristic initial 
value problem.

As for the initial condition inside the star, we have to give a data set not only for
metric perturbations but also for fluid and electromagnetic perturbations. 
As mentioned earlier, the initial fluid distribution $\bar{\beta}_0$ 
is given by equations (\ref{beta01})--(\ref{beta03}), while the initial 
distribution for the magnetic fields $b_2$, $b_3$ will be discussed later 
in \S \ref{sec:VI}. 
Finally, the momentarily static initial data for metric perturbations $\bar{\Pi}$ 
in region I defined by the conditions $\partial_{\eta}\bar{\Pi}=0$
and $\partial^2_{\eta}\bar{\Pi}=0$. 
The data sets at $\eta=0$ for $\partial_u \bar{\Pi}$ and
$\partial_v \bar{\Pi}$ are then defined via the relations 
$\partial_u \bar{\Pi} = -(\partial_{\chi}\bar{\Pi})/2$
and $\partial_v \bar{\Pi} = (\partial_{\chi}\bar{\Pi})/2$, respectively. 
The momentarily static distribution  of $\bar{\Pi}(\eta=0)$ due to the conditions
$\partial_{\eta}\bar{\Pi}=0$ and $\partial^2_{\eta}\bar{\Pi}=0$,  
is given as a regular solution of
\begin{align}
\partial_{\chi}^2 \bar{\Pi}+2(l+1)\frac{\cos\chi}{\sin\chi}\partial_{\chi}\bar{\Pi}-
\frac{(2l-1)(l+2)R(0)}{2R}\bar{\Pi}=\frac{16\pi}{R^l\sin^l\chi}
(\partial_{\chi}L_{\eta}-\partial_{\eta}L_{\chi}).
\label{ste}
\end{align}
The regular solutions of equation (\ref{ste}) must be smoothly connected to the static
exterior solution (\ref{static-solution}) through the stellar surface. 
This leads to a boundary condition for $\bar{\Pi}(\eta=0)$ at the surface of the 
star described  by the following equation
\begin{equation}
 \left(2l + 1 - \frac{rF_{l,r}}{F_l}\right) \bar{\Pi} + \tan \chi_0 \bar{\Pi}_{,\chi}
     + \frac{16\pi \beta \rho}{\left(R \sin \chi_0\right)^{l-1} \cos \chi_0}
     + \frac{16\pi \tilde{L}_t}{f\left(R \sin\chi_0 \right)^{l-1}} =0,
\end{equation}
where $F_l$ is the abbreviation for the hypergeometric function
$F_l(l-1,l+3,2l+2;2M/r)$. Finally, the regularity at the stellar center
requires that the function $\bar{\Pi}$ is analytic for $\chi \rightarrow 0$.

\subsection{Boundary Conditions}
\label{sec:IV-2}

The boundary conditions for the numerical integration are the regularity 
condition at the stellar center and that there are no incoming waves at the infinity. 
The regularity condition at the stellar center demans that
$\partial_\chi \bar{\Pi}=0$, which is reduced to
\begin{equation}
\frac{\partial \bar{\Pi}}{\partial u} = \frac{\partial \bar{\Pi}}{\partial v}.
\end{equation}
Finally, for the no incoming radiation condition at the infinity, we adopt the 
condition 
$\partial\tilde{\Phi}/\partial\tilde{u} = 0$ (see, e.g., \cite{Hamade1996}).

\subsection{Special Treatment of the Junction Conditions near the Event Horizon}

When the stellar surface reaches the event horizon, the junction conditions
discussed ealier in \S \ref{sec:III-4} cannot be used any more because the terms 
related to $f^{-1}$ diverge. 
Instead of these junction conditions, following \cite{Harada2003}, we impose 
the following junction conditions on
the null surface of $\tilde{v}=\tilde{v}_0$ in the vicinity of the 
point ${\cal H}$ in Fig.
\ref{fig-calculation-region}:
\begin{align}
 \Pi &= \Pi^{N_{\rm max}}
     + \frac{\Pi^{\rm EH} - \Pi^{N_{\rm max}}}{r^{\rm EH} - r^{N_{\rm max}}}\left(r - r^{N_{\rm max}}\right), \\
 \Pi^{\rm EH} &\equiv \Pi^{N_{\rm max}}
     + \frac{\Pi^{N_{\rm max}} - \Pi^{N_{\rm max}-1}}{r^{N_{\rm max}} - r^{N_{\rm max}-1}}
       \left(r^{\rm EH} - r^{N_{\rm max}}\right),
\end{align}
where $\Pi^{n}$ and $r^{n}$ are the values of $\Pi$ and $r$ on 
$\tilde{v}=\tilde{v}_0$ at $n$-th time steps, while $N_{\rm max}$ denotes the 
total number of time steps in region II, and $r^{\rm EH} = 2M$.

\section{Code tests}
\label{sec:V}

In order to check our numerical code, we have calculated the quadrupole gravitational 
radiation emitted during the collapse of a non-magnetized homogeneous dust sphere 
(perturbations of the Oppenheimer-Snyder solution), which has been already 
studied by several authors, e.g., \cite{Cunningham1978,Harada2003}. 
For the test we will consider the collapse of the homogeneous dust sphere which is initially at rest. 
Therefore, we have to provide the initial radius of the dust sphere to begin the 
numerical integration. 
Since the amount of gravitational radiation emitted during the collapse of a non-magnetized 
homogeneous dust sphere is the ``typical value'' with which we will compare the
energy emitted during the  collapse of a magnetized homogeneous dust 
sphere, we will briefly summarize these results.

In the present calculations, the number of the spatial grid points inside the star (region I) is chosen to be $N_{\chi}=1000$.
Using this number of grid points we manage to obtain numerical solutions and results with acceptable accuracy.  
In region III, the step-size for integration is determined by the relation
$\Delta \tilde{u} = (u_{\rm max} - u_0) / N_{\tilde{u}}$,
where $u_{\rm max} \equiv t_{\rm max} - r_{*{\rm ob}}$. 
Here, $t_{\rm max}$ is the expected maximum time for 
observation and $r_{*{\rm ob}}$ is the position of the observer described in
tortoise coordinate units. 
In this paper, we assume that $t_{\rm max} = 2000M$ while the fiducial observer 
is at $r_{{\rm ob}}=40M$.

\begin{center}
\begin{figure}[htbp]
\includegraphics[height=7cm]{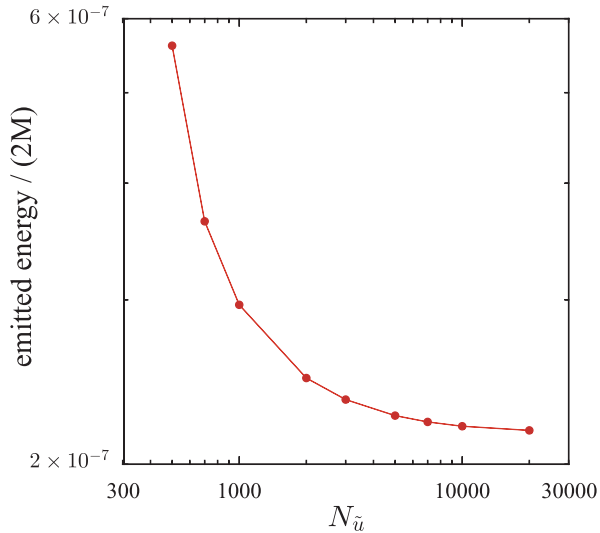}
\caption{The total radiated energie of quadrupole gravitational waves from the 
OS collapse characterized by the initial radius of the dust sphere $r_{s0} = 8M$ 
and $\bar{\beta}_0=const.$ velocity distribution, as function of the number of grid
points in region III.}
\label{fig-convergence}
\end{figure}
\end{center}

\begin{center}
\begin{figure}[htbp]
\includegraphics[height=7cm]{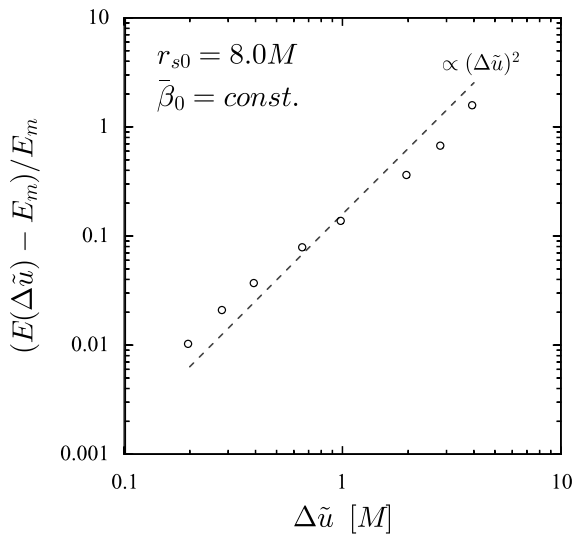}
\caption{
Convergence test of the numerical code. 
The vertical axis denotes the ``relative error",
that is the ratio of $E(\Delta\tilde{u}) - E_m$ over $E_m$,
where $E(\Delta\tilde{u})$ is the emitted energie for various of $\Delta\tilde{u}$ 
and $E_m$ is the energy for $N_{\tilde{u}}=20000$.
The dashed line is  $\propto (\Delta\tilde{u})^2$ suggesting  
that as $\Delta\tilde{u}$ becomes smaller
the ``relative error" reduces as $(\Delta\tilde{u})^2$.
}
\label{fig-error}
\end{figure}
\end{center}

As a first step we confirm the convergence of our numerical code. 
For this purpose, by varying the value of $N_{\tilde{u}}$, we calculate 
the total energy radiated in gravitational waves during the collapse
which will be characterized by the initial radius of the dust sphere, 
$r_{s0} = 8M$, and the $\bar{\beta}_0=const.$ velocity distribution. 
The radiated energy is estimated by integrating the luminosity of
gravitational waves $L_{GW,l}$ with respect to time. 
Here, the luminosity $L_{GW,l}$ of gravitational waves is defined by the relation
(see, e.g., \cite{Cunningham1978,Harada2003})
\begin{equation}
 L_{GW,l} = \frac{1}{16\pi}\frac{l(l+1)}{(l-1)(l+2)}(\tilde{\Phi}_{,\tilde{u}})^2.
\label{Lgw}
\end{equation}
The outcome of this calculation i.e. the energy emitted in gravitational waves during the collapse as a function of  $N_{\tilde{u}}$ is shown in Fig. \ref{fig-convergence} and tabulated in 
Table \ref{tab-test-data}. 
From Fig. \ref{fig-convergence} we conclude that the amount of the total energy
emitted during the collapse converges for $N_{\tilde{u}} \ge 10000$, 
thus we assumed in all numerical runs $N_{\tilde{u}}=10000$.
In Fig. \ref{fig-error}, we give the relative error
in the total emitted energy obtained by our numerical code as a function of 
$\Delta\tilde{u}$.
Here, the quantity $=(E(\Delta\tilde{u}) - E_m)/E_m$ stands for the relative error,
then $E(\Delta\tilde{u})$ denotes the total energy emitted  
for various  values of $\Delta\tilde{u}$ and $E_m$ is the energy for some
maximum value $N_{\tilde{u}}=20000$. 
It is obvious from Fig. \ref{fig-error} that our numerical code achieves 
second order accuracy.

%
%
%
%
\begin{table}[htbp]
\begin{center}
\leavevmode
\caption{The total radiated energies of quadrupole gravitational waves from the collapse
characterized by the initial radius of the dust sphere $r_{s0} = 8M$ and the
$\bar{\beta}_0=const.$ velocity distribution.}
\begin{tabular}{c c c c c}
\hline\hline
  & $N_{\tilde{u}}$ & $\Delta\tilde{u}$ & emitted energy / $(2M)$ & \\
\hline
  &  $500$   & $3.929M$    & $5.614\times 10^{-7}$  &  \\
  &  $700$   & $2.806M$    & $3.641\times 10^{-7}$  &  \\
  &  $1000$  & $1.964M$    & $2.964\times 10^{-7}$  &  \\
  &  $5000$  & $0.3929M$   & $2.255\times 10^{-7}$  &  \\
  &  $10000$ & $0.1964M$   & $2.197\times 10^{-7}$  &  \\
  &  $20000$ & $0.09822M$  & $2.175\times 10^{-7}$  &  \\
\hline\hline
\end{tabular}
\label{tab-test-data}
\end{center}
\end{table}
%
%

\begin{center}
\begin{figure}[htbp]
\includegraphics[height=7cm]{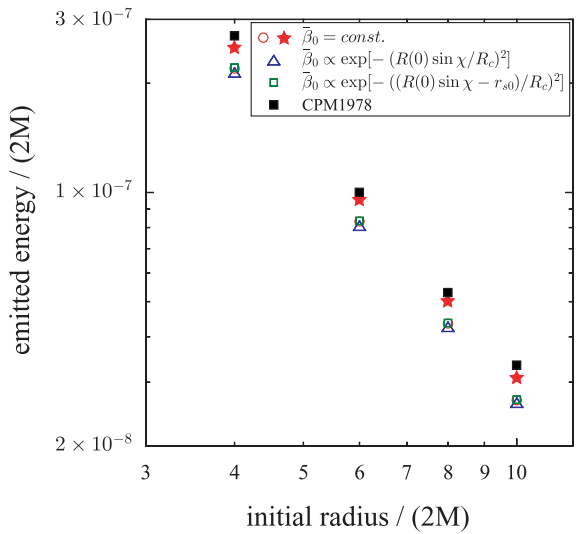}
\caption{
The energy emitted in gravitational waves  from the homogeneous 
dust collapse without magnetic field as a function of the initial stellar radius
where $r_{s0} = 8M$, $12M$, $16M$, and $20M$.
The filled-squares correspond to the results by \cite{Cunningham1978} 
while the rest correspond to our results. 
The different marks corresponds to different initial distribution  $\bar{\beta}_0$. 
The asterisks correspond to numerical results taken by a first order code.
}
\label{fig-emitted-energy}
\end{figure}
\end{center}

Next, let us compare the total energy emitted during the collapse as it has
been calculated by our numerical code with the results by
Cunningham, Price, and Moncrief \cite{Cunningham1978} (CPM1978). 
In Fig. \ref{fig-emitted-energy}, we show
the total energy emitted in gravitational waves during the collapse as  function 
of the initial radius of the dust sphere $r_{s0}$. 
In this figure, the results of CPM1978 are indicated by the filled-squares, while
the other symbols represent our results. 
The results for the initial distribution of the fluid velocity defined by
equation (\ref{beta01}) are indicated by circles, those defined by equation 
(\ref{beta02}) are indicated by triangles while for the distribution
defined by equation (\ref{beta03}) we used squares. 
Finally, the results obtained by using a numerical code with the first order 
accuracy, are indicated by the gray asterisks. 
This figure shows that there are small differences between our
results and those obtained by CPM1978. 
We however observe that the results of CPM1978 agree well with those with
the first order accuracy (compare the filled-squares with the asterisks). 
Therefore, we conclude that our results are in quite good agreement with the 
results of CPM1978.
%
%
%
%

\begin{figure}[htbp]
\begin{center}
\begin{tabular}{cc}
\includegraphics[scale=0.45]{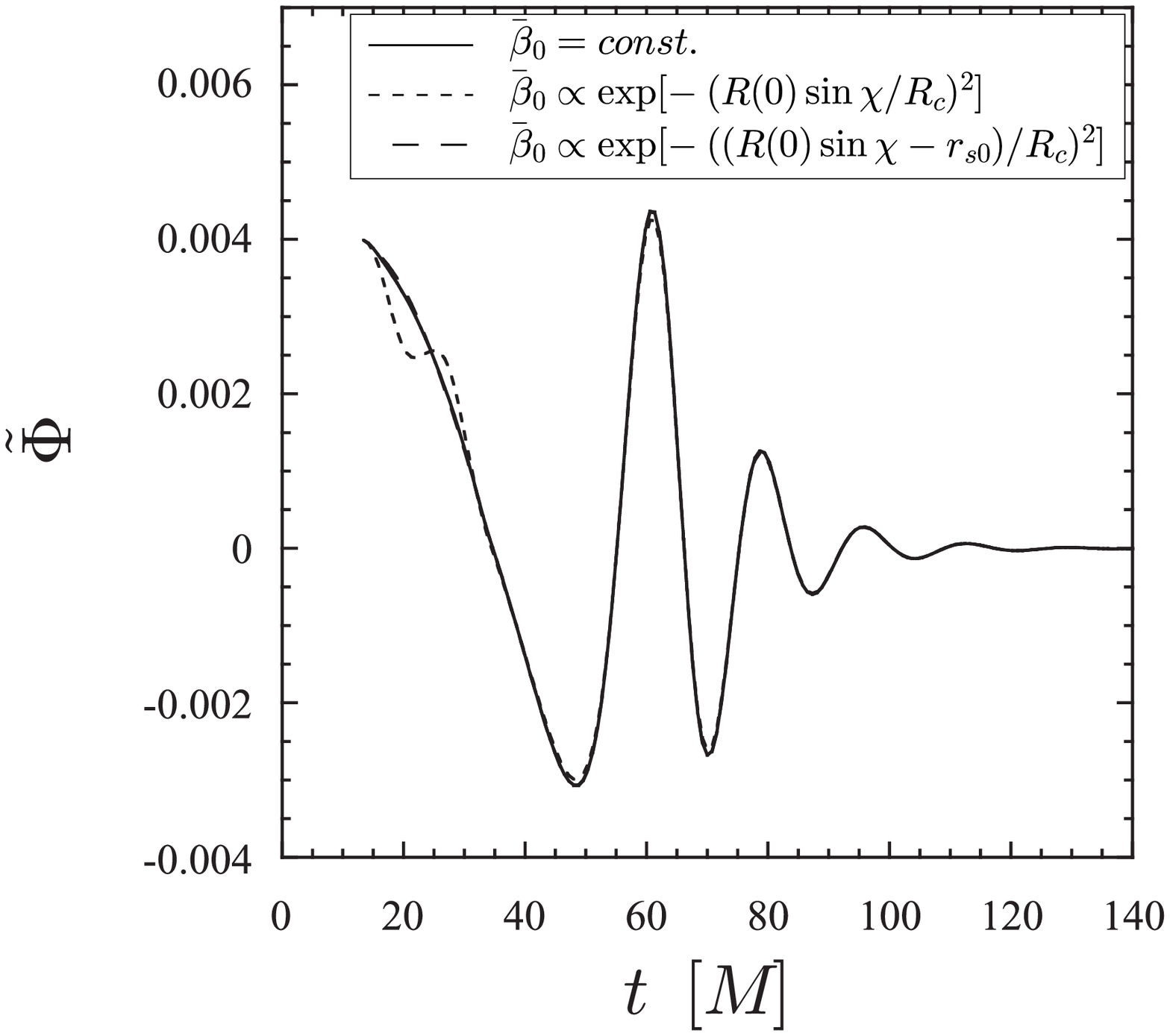} &
\includegraphics[scale=0.45]{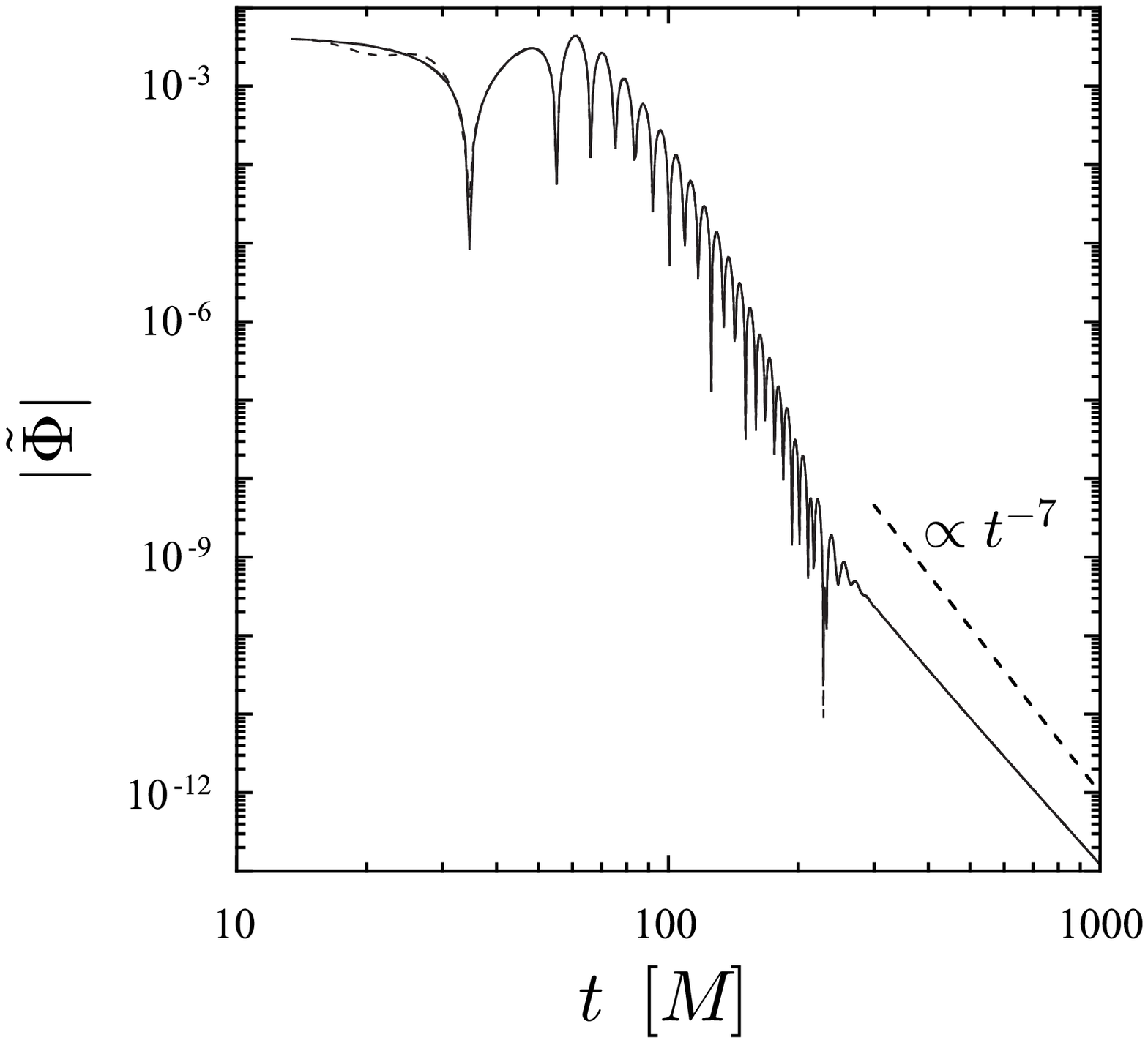} \\
\end{tabular}
\end{center}
\caption{The waveform of the quadrupole gravitational radiation emitted during 
the collapse of the non-magnetized homogeneous dust,  as function of  time.
The initial radius of the dust sphere is set to $r_{s0}=8M$
while the fiducial observer is set at $r=40M$. 
In the right panel the amplitudes of the gravitational waves are shown in a 
log-log plot and the late-time is compared with its theoretical value 
$t^{-(2l+3)}$.
Three initial distributions of the fluid velocity $\bar{\beta}_0$ were adopted 
and one can hardly trace their influence on the waveform.}
\label{fig-test}
\end{figure}

In Fig. \ref{fig-test}, we show the waveforms of the quadrupole gravitational 
radiation from the collapse with an initial radius $r_{s0}=8M$ and for three 
different initial distributions of the fluid velocity $\bar{\beta}_0$
defined by equations (\ref{beta01}), (\ref{beta02}), and (\ref{beta03}). 
In this figure, the left panel displays the 
waveforms as functions of the time, while the right panel displays the 
absolute values of the amplitudes as functions of the time in a log-log plot. 
Fig. \ref{fig-test} shows that the first part of the waveform is characterized by
the quasinormal ringing while at the late times follows a power-low tail, 
as found in \cite{Cunningham1978}.
(For a review of quasinormal modes for compact objects, see, e.g. \cite{ks1992}.)
From the waveforms, we estimate the frequency of the fundamental quasinormal 
mode to be $2M\omega = 0.746 + 0.179i$, which agrees very well with the quasinormal 
mode frequency estimated by Chandrasekhar and Detweiler \cite{Chandrasekhar1975}. 
As for the late-time tail of the gravitational waves, in the right
panel of Fig. \ref{fig-test}, we find that the amplitude decays as $t^{-7}$ at 
late times, which is in good agreement with the analytical estimate of 
Price \cite{Price1972}, that is $t^{-(2l+3)}$. 
The accuracy in the estimates of the quasinormal mode frequency and the 
late-time tail therefore suggest that our numerical code is accurate and 
reproduces all previously known results.

Besides the tests of the code  the following basic 
properties of the gravitational radiation emitted
during the collapse of the non-magnetized homogeneous dust 
should be emphasized. First, as it shown in Fig.
\ref{fig-emitted-energy}, the total radiated energy 
does not critically depend on the distribution of
$\bar{\beta}_0$, this has been also observed in CPM1978 and 
second, as shown in Fig. \ref{fig-test}, that small
modulations appear just after the onset of the collapse only for the case of
$\bar{\beta}_0\propto \exp[-(R(0)\sin\chi/R_c)^2]$.
%
%
%
%
%
%
%

\section{Gravitational radiation from the collapse of the homogeneous magnetized 
dust sphere}
\label{sec:VI}

\subsection{Initial distribution of the magnetic field and magnetic effects on
the gravitational radiation}
\label{sec:VI-0}

For the calculation of the gravitational waves emitted during the collapse of 
a magnetized dust sphere, one needs to provide the initial 
distribution (profile) of the magnetic fields, i.e., to set up the functional forms of 
$b_2$ and $b_3$ on the hypersurface $\eta=0$.
In practice, one has the freedom in choosing the initial distribution of 
$b_2$ and $b_3$  as the following two conditions are satisfied: 
(i) the regularity condition at the stellar center, 
(ii) the junction condition (\ref{magcon}) at the stellar surface. 
Since here we made the assumption that the magnetic field is confined 
inside the star, the second condition is reduced to $b_2(\chi_0)=0$. 
In this work we assume a dipole magnetic field, and for this specific geometry
it is convenient to introduce two new quantities for the description of the
magnetic field,  $\bar{b}_2$ and
$\bar{b}_3$, defined as
\begin{align}
 b_2(\chi) &= (R \sin\chi)^2 \bar{b}_2(\eta,\chi)\,, \\
 b_3(\chi) &= (R \sin\chi)^3 \bar{b}_3(\eta,\chi)\, .
\end{align}
If the new variables $\bar{b}_2(\eta,\chi)$ and $\bar{b}_3(\eta,\chi)$ are 
analytic at $\chi=0$, the regularity condition at the stellar center for the 
magnetic field is automatically satisfied. 
Then for quadrupole perturbations ($l=2$), the source term in the wave equation, 
$S(\eta,\chi)$, and the perturbation for the four-velocity 
$\bar{\beta}(\eta,\chi)$ are given by the following expressions:
\begin{align}
 S(\eta,\chi) &= 4\pi R^2 \rho\left\{3\bar{\beta}\cos\chi + (\partial_{\chi}\bar{\beta}) \sin\chi\right\}
    - \frac{R(0)\sin\eta}{3}\left(\frac{R(0)}{R}\right)^5 {\cal B}_2 {\cal B}_3
   \bar{b}_{20}\bar{b}_{30} \sin\chi,
    \label{source-M}\\
 \bar{\beta}(\eta,\chi) &= \frac{2\eta R^2}{9R(0)}\left(\frac{R(0)}{R}\right)^5 {\cal B}_2{\cal B}_3
       \left\{\bar{b}_{20} \bar{b}_{30} \tan\chi
     + \bar{b}_{20} (\partial_{\chi}\bar{b}_{30}) - (\partial_{\chi}\bar{b}_{20}) \bar{b}_{30}\right\}
     + \left(\frac{R(0)}{R}\right)^3 \bar{\beta}_0(\chi), \label{beta-M}
\end{align}
where $\bar{b}_{20}$, $\bar{b}_{30}$ are dimensionless functions of $\chi$, defined as
\begin{align}
 \bar{b}_2(\eta,\chi) &= \left(\frac{R(0)}{R}\right)^2 {\cal B}_2 \, \bar{b}_{20}(\chi), \\
 \bar{b}_3(\eta,\chi) &= \left(\frac{R(0)}{R}\right)^3 {\cal B}_3 \, \bar{b}_{30}(\chi).
\end{align}
Here ${\cal B}_2$ and ${\cal B}_3$ are arbitrary constants related to the 
strength of the magnetic field.
It should be emphasized that, as shown in equations (\ref{source-M}) and (\ref{beta-M}), all terms related to the magnetic fields in the source term $S(\eta,\chi)$ vanish when $\eta=0$. 
In other words, the magnetic field does not affect the momentarily static initial data for metric perturbations. 
The geometry of the magnetic fields when the collapse sets in is practically unknown. 
Based on this freedom we adopted the following two types for the initial distribution of the magnetic field:
\begin{align}
\mbox{(I)}\ :&\ \  \bar{b}_{20}(\chi) = 1 - 2\left(\frac{\chi}{\chi_0}\right)^2
     + \left(\frac{\chi}{\chi_0}\right)^4,\ \
      \bar{b}_{30}(\chi) = 1 - \left(\frac{\chi}{\chi_0}\right)^4, \label{MF1} \\
\mbox{(II)}\ :&\ \  \bar{b}_{20}(\chi) = \bar{b}_{30}(\chi)
     = 4\left[\left(\frac{\chi}{\chi_0}\right)^2 - \left(\frac{\chi}{\chi_0}\right)^4\right]. \label{MF2}
\end{align}
Note that the maximum value of $\bar{b}_{20}$ and $\bar{b}_{30}$ are chosen
to be one in the interval $0\le \chi \le \chi_0$.
For the first profile function (I) the magnetic field is stronger in the center of the sphere while  for profile function (II) the field becomes stronger in the outer region.

It is more convenient to explore the effects of the magnetic field on the efficiency of the gravitational radiation emission during the collapse, if one introduces the dimensionless parameter $\alpha$, defined as
\begin{align}
\alpha=\frac{R(0){\cal B}_2{\cal B}_3}{{\cal U}_1}\,.
\end{align}
Then the source term in the wave equation, $S(\eta,\chi)$, can be split as follows
\begin{align}
S=S^{(\beta)} + \alpha S^{(B)}\,,
\label{smag}
\end{align}
where
\begin{eqnarray}
S^{(\beta)}&=&4\pi R(0)^2 \rho\left(\frac{R(0)}{R}\right)
\left\{3\bar{\beta}_0\cos\chi + (\partial_{\chi}\bar{\beta}_0) \sin\chi\right\}\,, \\
S^{(B)}&=&{\cal U}_1\left(\frac{R(0)}{R}\right)^5\left[\frac{8\pi}{9R(0)^2}R^4\rho\eta 
\frac{1}{\sin^2\chi}
\partial_{\chi}\biggl\{\biggl(\bar{b}_{20}\bar{b}_{30}\tan\chi+\bar{b}_{20}(\partial_{\chi}\bar{b}_{30})-
(\partial_{\chi}\bar{b}_{20})\bar{b}_{30}\biggr)\sin^3\chi\biggr\} \right. \nonumber \\
&& \left.\quad\quad\quad\quad\quad\quad\quad - \frac{\sin\eta}{3}\bar{b}_{20}\bar{b}_{30}\sin\chi
\right]\,.
\end{eqnarray}
Note that $S^{(\beta)}$ can be attributed to the incompressible fluid flow 
while $S^{(B)}$ to the Maxwell stress in the magnetized dust sphere. 
The splitting of the source term introduced with equation (\ref{smag}) suggests 
that any solution of the wave equations (\ref{interior_GW1}) and 
(\ref{exterior_GW}) can be expressed as a linear superposition of the two 
solutions, $\tilde{\Phi}^{(\beta)}$ and $\tilde{\Phi}^{(B)}$, that are 
independent of $\alpha$, i.e.
\begin{equation}
 \tilde{\Phi} = \tilde{\Phi}^{(\beta)} + \alpha \tilde{\Phi}^{(B)}\,, \label{eq:decompose}
\end{equation}
where $\tilde{\Phi}^{(\beta)}$ is the solution in the absense of magnetic field
($\alpha=0$) and $\tilde{\Phi}^{(B)}$ is the solution for the case of $S=S^{(B)}$
i.e. when the gravitational field is initially stationary.
Since we assume that the initial profile of the gravitational perturbations 
does not dependent on the existence of magnetic fields, the initial values 
of $\tilde{\Phi}^{(B)}$ were set to zero.

These assumtions, i.e. the splitting introduced by equations (\ref{smag}) and 
(\ref{eq:decompose}), suggest modifications in the form of equation (\ref{Lgw})
describing the luminosity in gravitational waves which gets the following form:
\begin{equation}
 L_{GW,l} = \frac{1}{16\pi}\frac{l(l+1)}{(l-1)(l+2)}\{(\tilde{\Phi}^{(\beta)}_{,\tilde{u}})^2 +
\alpha^2(\tilde{\Phi}^{(B)}_{,\tilde{u}})^2+
2\alpha (\tilde{\Phi}^{(\beta)}_{,\tilde{u}})(\tilde{\Phi}^{(B)}_{,\tilde{u}})
\}\,.
\label{Lgw2}
\end{equation}
While the radiated energy in gravitational waves during the collapse
is then given by following relation
\begin{equation}
E_{GW,l}=E_{GW,l}^{(\beta)}+\alpha^2 E_{GW,l}^{(B)}+2\alpha C_{GW,l} \,,
\label{Egw}
\end{equation}
where $E_{GW,l}^{(\beta)}$ and $E_{GW,l}^{(B)}$ stand for the total radiated 
energies associated with the solutions $\tilde{\Phi}^{(\beta)}$ and 
$\tilde{\Phi}^{(B)}$, respectively, and $C_{GW,l}$ is an integral quantity 
defined by the product of $\tilde{\Phi}^{(\beta)}$ and $\tilde{\Phi}^{(B)}$. 
Their detailed form is:
\begin{eqnarray}
E_{GW,l}^{(\beta)}&=&\frac{1}{16\pi}\frac{l(l+1)}{(l-1)(l+2)}
\int_{\tilde{v}=\tilde{v}_{max}}(\tilde{\Phi}^{(\beta)}_{,\tilde{u}})^2 d\tilde{u}
\,, \\
E_{GW,l}^{(B)}&=&\frac{1}{16\pi}\frac{l(l+1)}{(l-1)(l+2)}
\int_{\tilde{v}=\tilde{v}_{max}}(\tilde{\Phi}^{(B)}_{,\tilde{u}})^2 d\tilde{u}
\,,\\
C_{GW,l}&=&\frac{1}{16\pi}\frac{l(l+1)}{(l-1)(l+2)}
\int_{\tilde{v}=\tilde{v}_{max}}(\tilde{\Phi}^{(\beta)}_{,\tilde{u}})(\tilde{\Phi}^{(B)}_{,\tilde{u}})d\tilde{u}
\,.
\end{eqnarray}
where $\tilde{v}_{max}$ is the maximum value of $\tilde{v}$.

It is worth mentioning the following issue emerging from the study of 
equation (\ref{Egw}). It can be easily proved that since $E_{GW,l}$ is a quadratic 
function of $\alpha$ and $E_{GW,l}^{(B)}>0$ then by definition, $E_{GW,l}$ gets 
its minimum value, $E_{GW,l}^{(min)}=E_{GW,l}^{(\beta)} - {C_{GW,l}}^2/E_{GW,l}^{(B)}$,
for $\alpha=\alpha_c$, where
$\alpha_c=-C_{GW,l}/E_{GW,l}^{(B)}$. 
In other words, phase cancellation between the two components 
$\tilde{\Phi}^{(\beta)}$ and $\tilde{\Phi}^{(B)}$ of the gravitational perturbations
become maximal around $\alpha=\alpha_c$.
As discussed earlier in \S \ref{sec:V}, $E_{GW,l}^{(\beta)}$ does not highly 
depend on the initial profile of the fluid velocity. 
Thus, the total radiated energy $E_{GW,l}$ practically depends on $\alpha$ and
the initial radius of the dust sphere, $r_{s0}$, but not on $\bar{\beta}_0$.

\subsection{Numerical results for the initial magnetic field profile (I)}
\label{sec:VI-1}


The numerical study for the influence of the magnetic field on the gravitational  wave output during the OS collapse, we have mentioned earlier, that we used two quite different initial profiles for the magnetic field.
These two profiles described by the equations  (\ref{MF1}) and (\ref{MF2}) and represent magnetic fields which have their maximum either at the stellar center or near the surface.

\begin{figure}[htbp]
\begin{center}
\includegraphics[height=7cm]{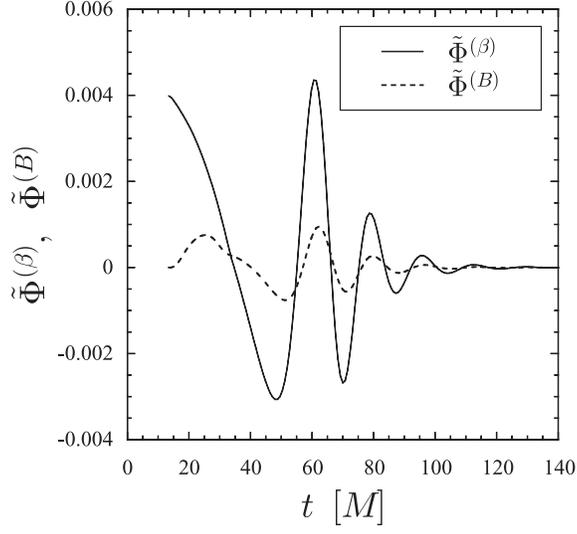}
\end{center}
\caption{
The amplitude of quadrupole gravitational 
waves emitted during the collapse. The continuous line stands for the function 
$\tilde{\Phi}^{(\beta)}$ and the dashed for $\tilde{\Phi}^{(B)}$. 
The efficiancy of the collapse depends on the initial radius of the 
dust sphere $r_{s0}=8M$, 
the initial  profile of the fluid velocity $\bar{\beta}_0=const.$, and
the initial pofile of the magnetic field, here is the case  (I).
The waveforms are monitored at $r=40M$.
}
\label{fig:decompose}
\end{figure}

We first consider the case (I), given by equation (\ref{MF1}). 
As discussed in the previous subsection, the fundamental quantities for 
estimating the gravitational radiation from the collapse of the magnetized 
homogeneous dust sphere are $\tilde{\Phi}^{(\beta)}$ and $\tilde{\Phi}^{(B)}$.
It is obvious that we can obtain solutions of equations (\ref{interior_GW1}) 
and (\ref{exterior_GW}) for any value of $\alpha$ in terms of 
$\tilde{\Phi}^{(\beta)}$ and $\tilde{\Phi}^{(B)}$ through the relationship 
(\ref{eq:decompose}). 
In Fig. \ref{fig:decompose}, we show the waveforms 
$\tilde{\Phi}^{(\beta)}$ and $\tilde{\Phi}^{(B)}$ of the quadrupole 
gravitational radiation from the collapse characterized by
an initial radius $r_{s0}=8M$ and 
an initial distribution of the fluid velocity $\bar{\beta}_0=const.$
We observe in Fig. \ref{fig:decompose} that the two waveforms,
$\tilde{\Phi}^{(\beta)}$ and $\tilde{\Phi}^{(B)}$, are almost in phase.
Therefore, we expect the following properties of the gravitational wave amplitude: 
(1) for $\alpha >0$, as $\alpha$ increases, the gravitational wave amplitude 
$\tilde{\Phi}$ increases monotonically and the phase of $\tilde{\Phi}$ does 
not change, 
(2) for $\alpha <0$, as $\alpha$ decreases, the amplitude and phase of the 
gravitational wave amplitude $\tilde{\Phi}$ show a more involved behavior
due to the phase cancellation between the two gravitational wave amplitudes
$\tilde{\Phi}^{(\beta)}$ and $\tilde{\Phi}^{(B)}$. 
It has been also found that the amplitude of the emitted gravitational waves is 
almost independent from the functional form of $\bar{\beta}_0$ in agreement with our
previous results for the non-magnetized collapse (see Fig. \ref{fig-test}).
%
%
%

In Figure 7 we show the waveforms for a number of positive and negative
values of $\alpha$ i.e. for $\alpha=-9$, $-6$, $-3$, $3$, $6$, and $9$. 
We also show, in every panel, the shape of the waveform in the absense of 
magnetic field ($\alpha=0$) with a bold line. 
This figure verifies the previosly refered  theoretical estimations i.e. 
that for $\alpha >0$, the phase of the various waveforms is almost the same as 
that  of $\alpha=0$ and the amplitude becomes larger as $\alpha$ increases. 
The amplitude of the waveforms is considerably smaller for   $\alpha < 0$, 
due to the phase cancellation effects, and decreases for smaller values of  
$|\alpha|$ on the other hand the phase shift is significant for large
values of  $|\alpha|$. 
Finally, one can easily observe that the dependence of the amplitude on the initial profile of $\bar{\beta}_0$  is very weak. 
Moreover, we  observe another effect related to the magnetic field i.e.  there is a wave packet the actual   quasinormal ringing. The appearances of this wave packet can be only attributed to the 
functional form of $\tilde{\Phi}^{(B)}$.

%
\begin{figure}[htbp]
\begin{center}
\begin{tabular}{ccc}
\includegraphics[scale=0.4]{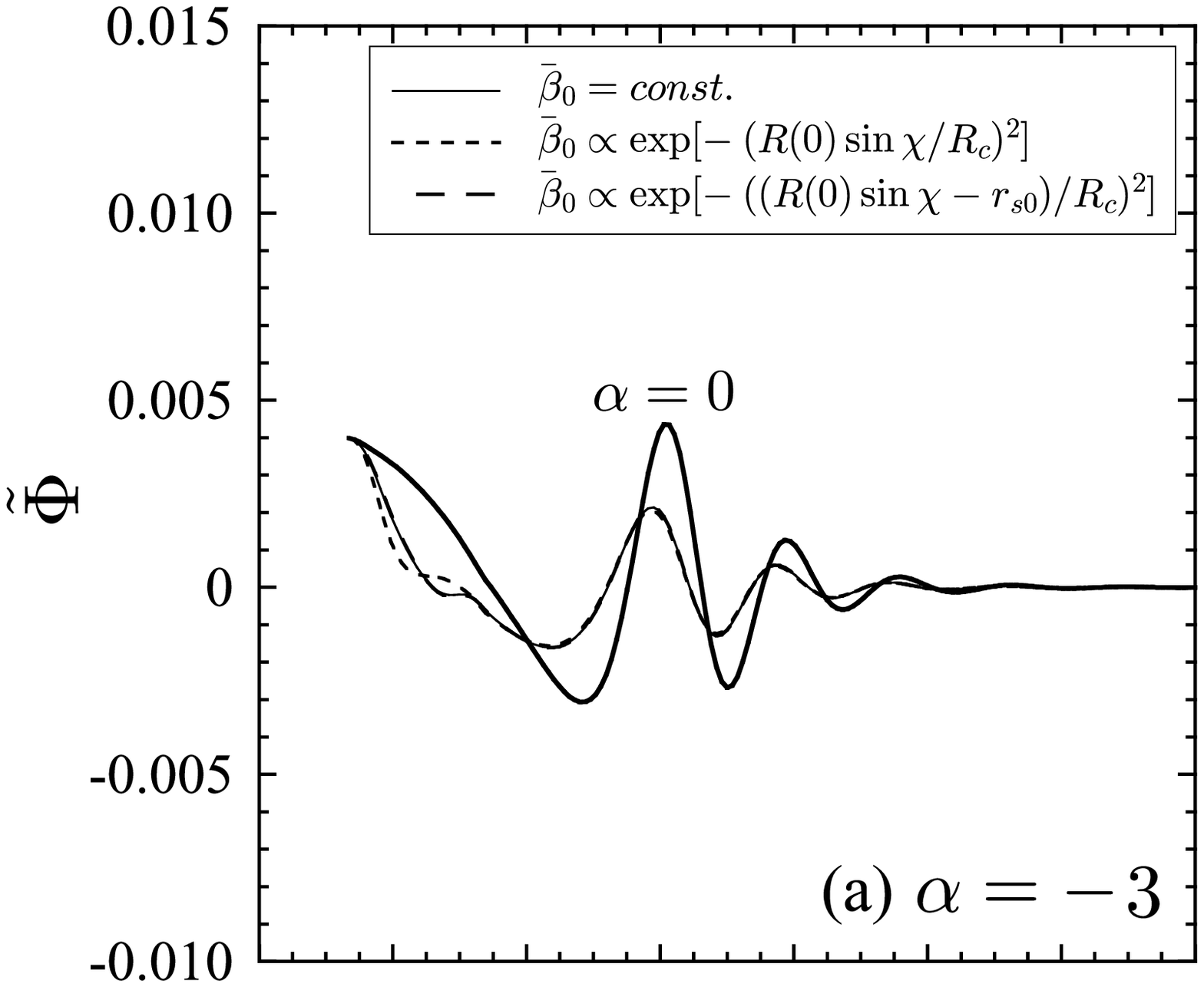} &
\includegraphics[scale=0.4]{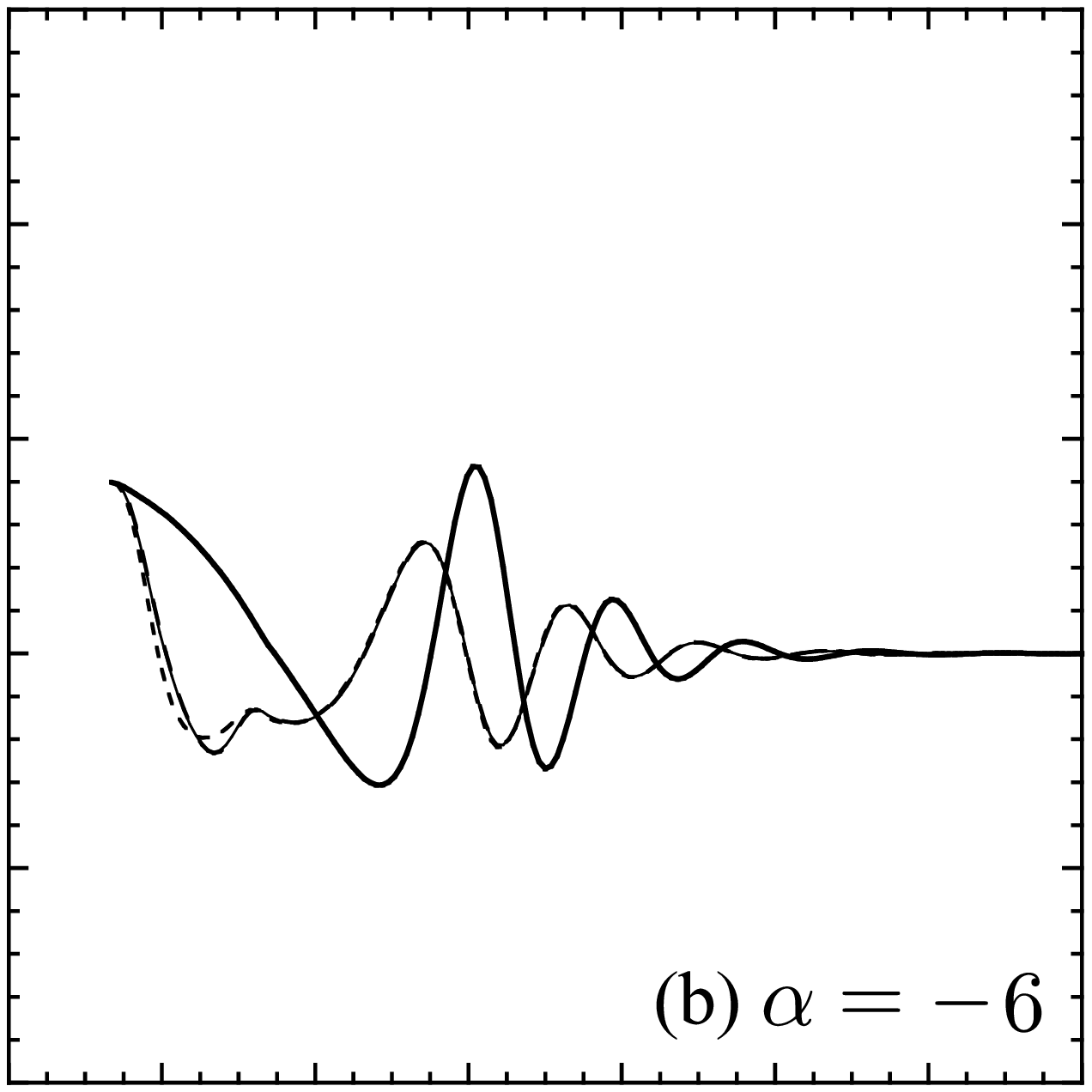} &
\includegraphics[scale=0.4]{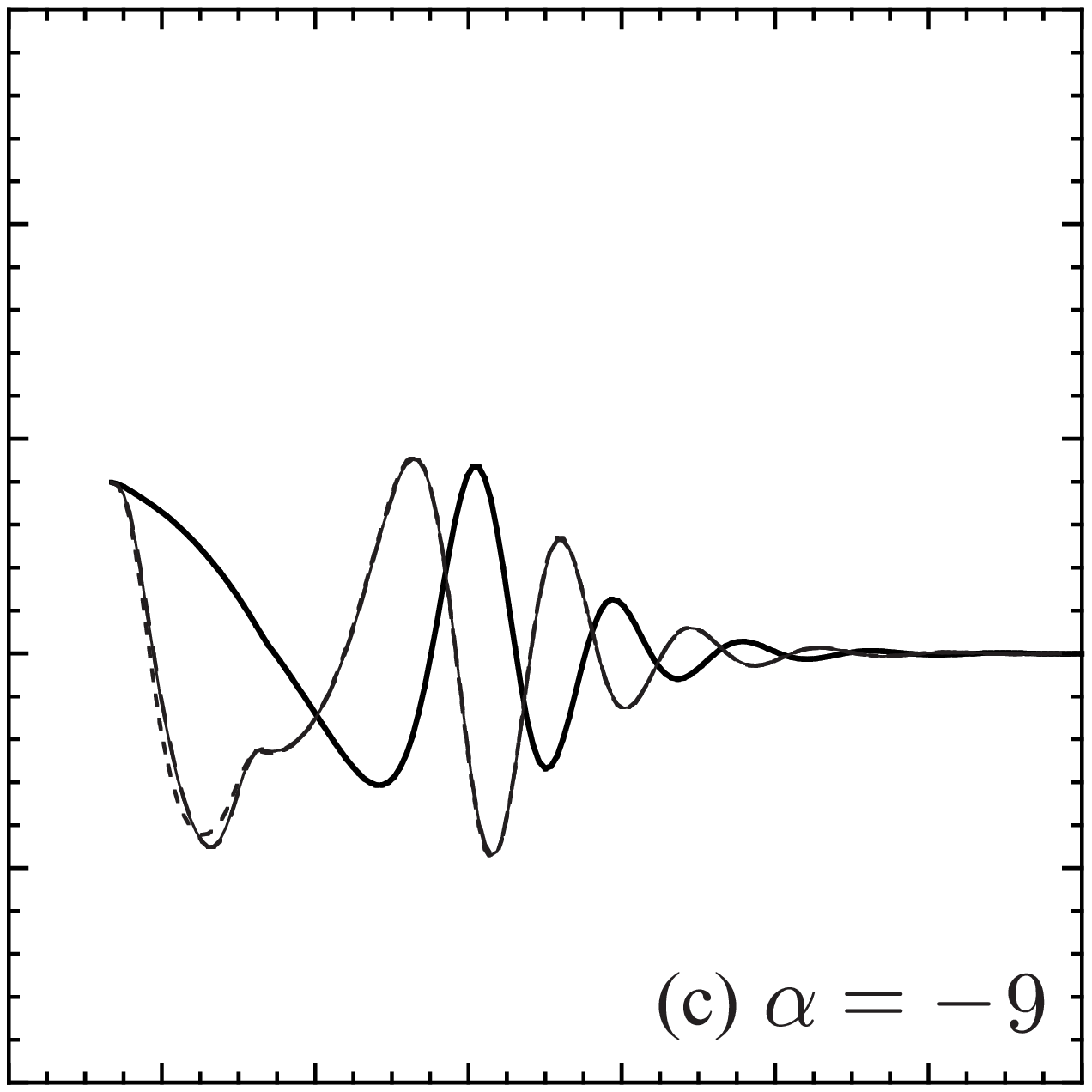} \\
\includegraphics[scale=0.4]{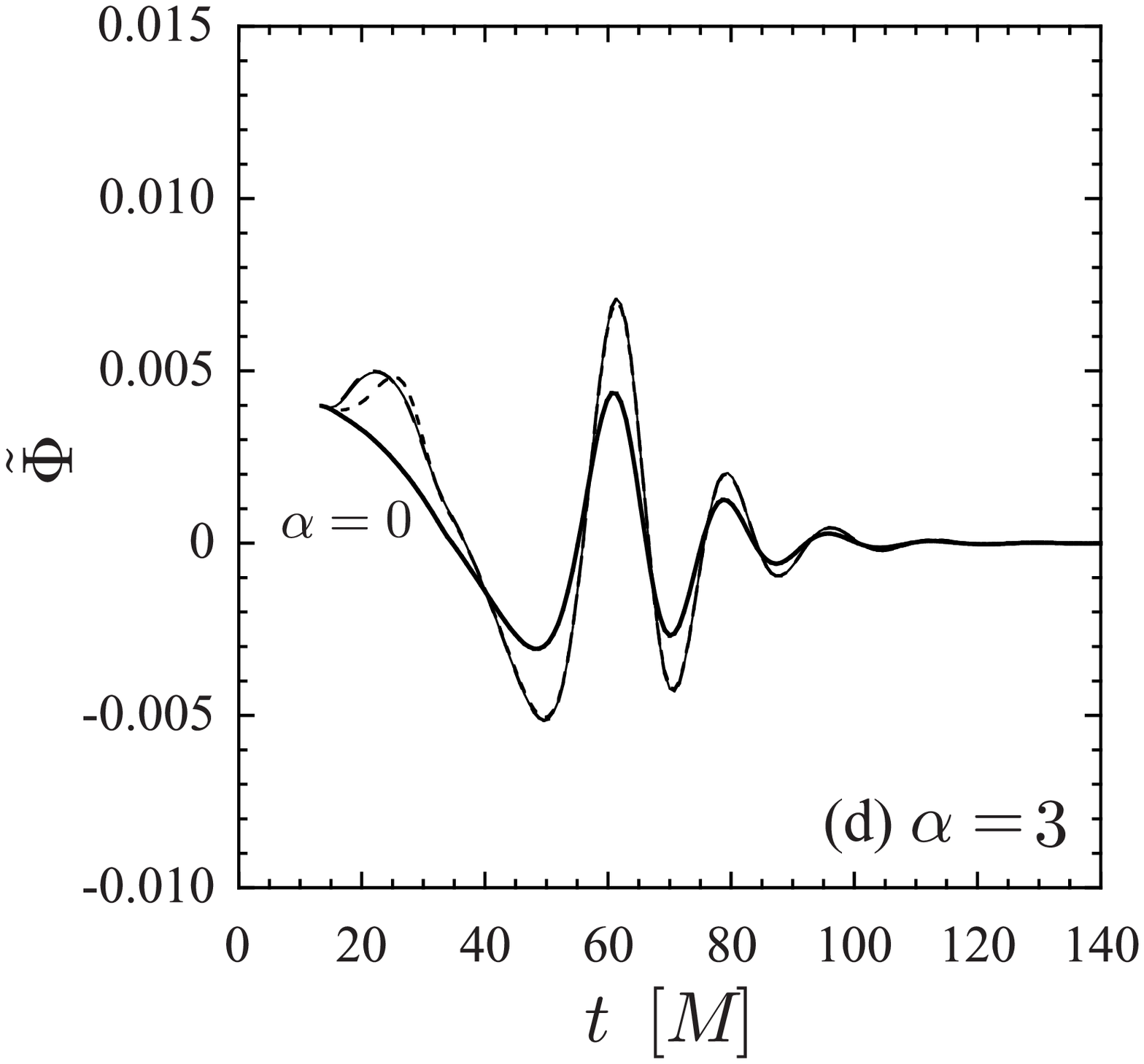} &
\includegraphics[scale=0.4]{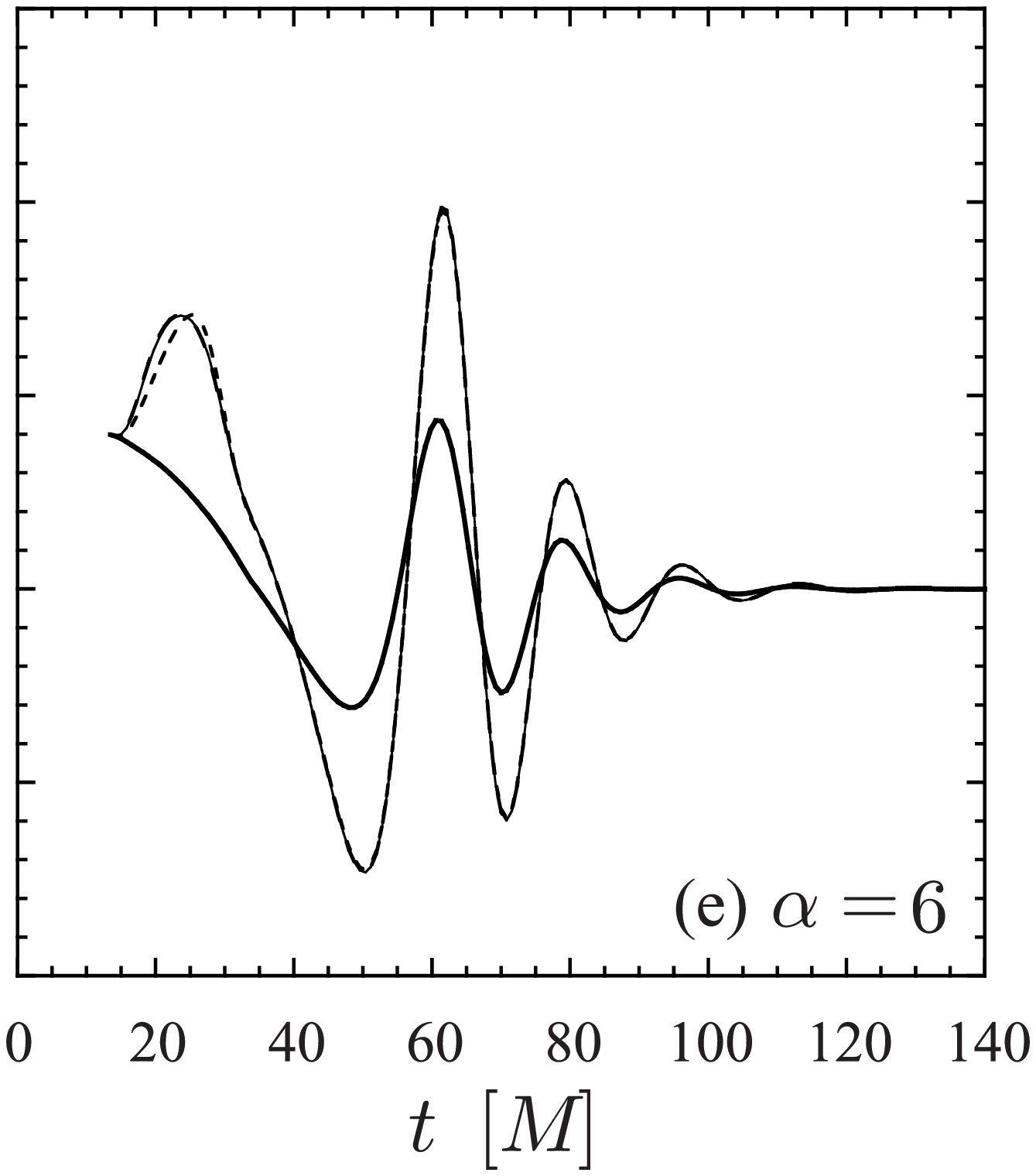} &
\includegraphics[scale=0.4]{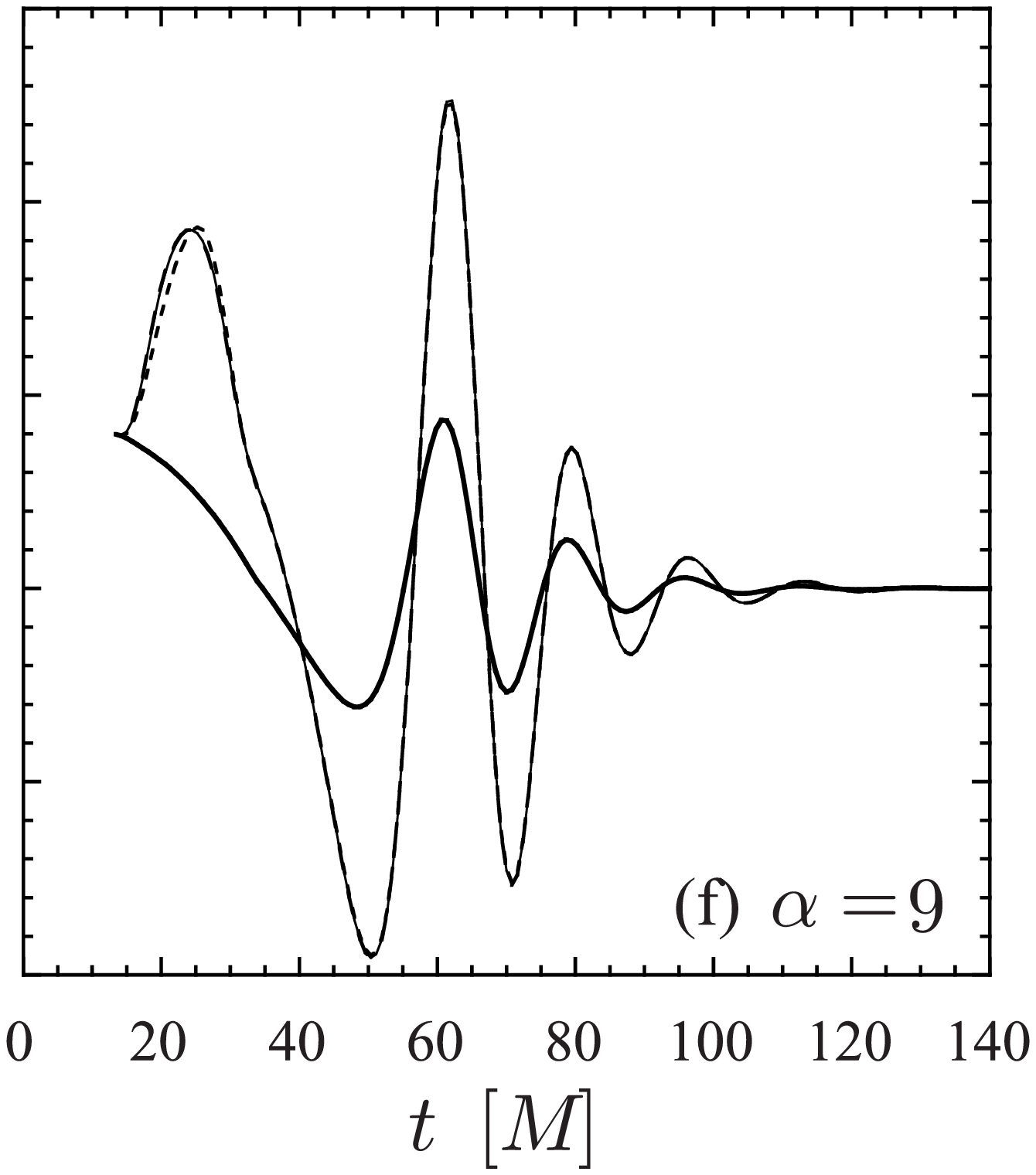} \\
\end{tabular}
\end{center}
\caption{\label{fig-waveform1}
Waveforms for gravitational radiation emitted during the collapse of   
of a magnetized homogeneous dust sphere with initial radius $r_{s0}=8M$ and
initial distribution of the fluid velocity $\bar{\beta}_0=const.$. 
The gravitational waveforms are monitored at $r=40M$. 
The thick continous line corresponds to the waveform of the non-magnetized 
collapse ($\alpha=0$).
}
\end{figure}
\begin{center}
\begin{figure}[htbp]
\includegraphics[height=7cm]{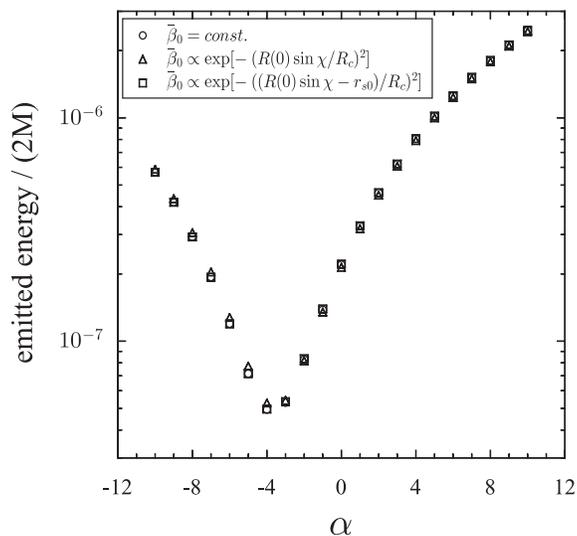}
\caption{
The total energy emitted in gravitational waves ($l=2$) during the homogeneous 
dust collapse with magnetic field as function of  $\alpha$.
The horizontal axis denotes the value of the parameter $\alpha$ representing 
the strength of the magnetic field, while the initial radius of the sphere 
assumed to be $r_{s0} = 8M$.
}
\label{fig-alpha-E-R08}
\end{figure}
\end{center}

As it has been argued earlier, the total energy emitted  in gravitational 
waves during the collapse of a magnetized dust sphere for any value of $\alpha$ 
can be calculated as a proper combination of the quantities 
$E_{GW,l}^{(\beta)}$, $E_{GW,l}^{(B)}$, and $C_{GW,l}$ via equation (\ref{Egw}).
The values of $E_{GW,2}^{(\beta)}$, $E_{GW,2}^{(B)}$, and $C_{GW,2}$ with
$\bar{\beta}_0=const.$ for four different initial radius of the dust sphere 
($r_{s0}=8M$, $12M$, $16M$, $20M$) are summarized in Table \ref{tab-energy}.
By studying equation (\ref{Egw}) we observe that the total energy emitted in gravitational waves has a minimum for collapsing models with $\alpha=\alpha_c <0$.
Since $E_{GW,2}^{(\beta)}$'s are almost independent from the initial distribution 
of the fluid velocity $\bar{\beta}_0$, as shown in Fig. \ref{fig-emitted-energy}, 
the total emitted energy hardly depends on $\bar{\beta}_0$. These features can
be seen in Fig. \ref{fig-alpha-E-R08} where the total energy emitted 
in gravitational waves from the collapse is studyed as function of $\alpha$

\begin{center}
\begin{figure}[htbp]
\includegraphics[height=7cm]{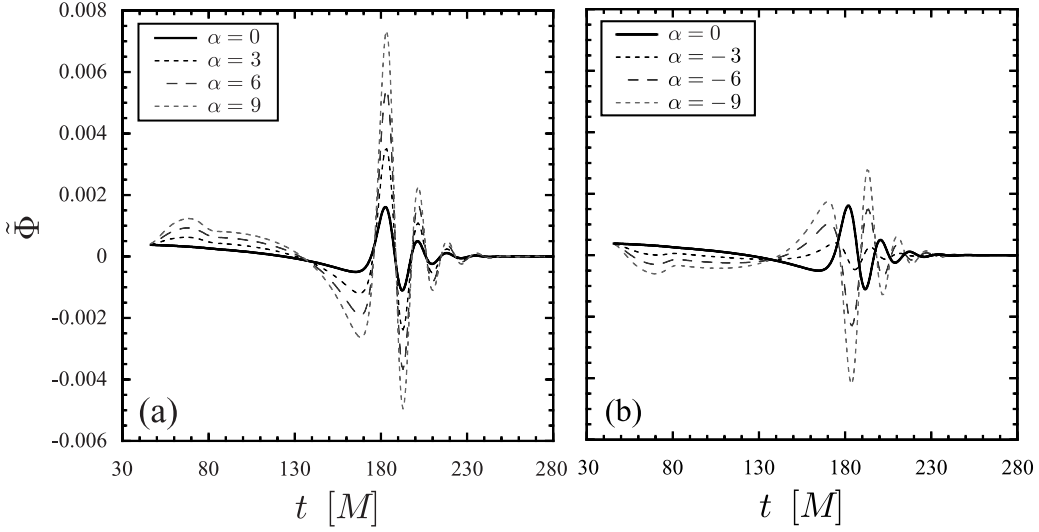}
\caption{
Waveform of gravitational waves for $l=2$ for (a) $\alpha \ge 0$ and (b) $\alpha \le 0$.
The initial radius is $r_{s0}=20M$ while the observer is set at $r=40M$ 
and we also assume that 
$\bar{\beta}_0=const.$
}
\label{fig-waveformR20}
\end{figure}
\end{center}

The most important conclusion that can be drawn from  this figure is that  the total energy emitted in gravitational waves from the magnetized dust collapse can be about eleven times higher than the energy of the non-magnetized collapse for $\alpha =10$ and about five times smaller for  $\alpha=\alpha_c$.
Thus, the effect of the magnetic field in the gravitational wave outcome during the collapse can be significant and might improve the possibility of detecting gravitational waves from this type of sources.
%
%
\begin{table}[htbp]
\begin{center}
\leavevmode
\caption{The values of $E_{GW,l}^{(\beta)}$, $E_{GW,l}^{(B)}$, and $C_{GW,l}$ for
quadrupole gravitational waves ($l=2$) emitted during the dust collapse
for four values of the initial radius of the dust sphere ($r_{s0}=8M$, $12M$, $16M$, and
$20M$) while we assumed that $\bar{\beta}_0 = const.$
The value, $\alpha_c \equiv -C_{GW,2}/E_{GW,2}^{(B)}$, i.e. the minimum of the 
emitted energy $E_{GW,2}$  is also shown.
}
\begin{tabular}{cc|ccccccccc}
\hline\hline
  & & & $r_{s0}=8M$ & & $r_{s0}=12M$ & & $r_{s0}=16M$ & & $r_{s0}=20M$ & \\
\hline
  & $E_{GW,2}^{(\beta)}/(2M)$ & & $2.197\times 10^{-7}$ & & $8.319\times 10^{-8}$ &
& $4.354\times 10^{-8}$ & & $2.676\times 10^{-8}$ & \\
  & $E_{GW,2}^{(B)}/(2M)$     & & $1.296\times 10^{-8}$ & & $7.527\times 10^{-9}$ &
& $5.557\times 10^{-9}$ & & $4.472\times 10^{-9}$ & \\
  & $C_{GW,2}/(2M)$           & & $4.722\times 10^{-8}$ & & $2.366\times 10^{-8}$ &
& $1.501\times 10^{-8}$ & & $1.066\times 10^{-8}$ & \\
& $\alpha_c$            & & $-3.64$ & & $-3.14$ & & $-2.70$ & & $-2.38$ & \\
\hline\hline
\end{tabular}
\label{tab-energy}
\end{center}
\end{table}
%
%
%
%
Next we examine the dependence of the total energy emitted during the collapse
on  the initial radius $r_{s0}$ of the dust sphere assuming again that 
the initial profile of the fluid velocity is  $\bar{\beta}_0=const.$ 
In Fig. \ref{fig-waveformR20}, the waveforms from the collapse with 
an initial radius $r_{s0}=20M$ are shown.
From this figure, can be easily seen  that the effect of the magnetic field becomes 
more pronounced as the initial stellar radius increases while the basic 
properties (dependence on $\alpha$ and phase shift) are similar to those of 
the $r_{s0}=8M$ case.
The influence of the magnetic field in the gravitational wave output increases 
with increasing radius, a natural explanation is that the longer the collapse 
lasts  the longer the magnetic field will influence the dynamics of the collapsing dust.
Actually, if the collapsing sphere has an initial raidus $r_{s0}=8M$ then 
it takes $t\sim 50M$ until one observes the first peak of the quasinormal ringing,
while it takes $t\sim140M$ for the $r_{s0}=20M$ model.
Additionally, we can observe that a wave packet before the quasinormal ringing
is suppressed, compare with Fig. \ref{fig:decompose}.

\begin{center}
\begin{figure}[htbp]
\includegraphics[height=7cm]{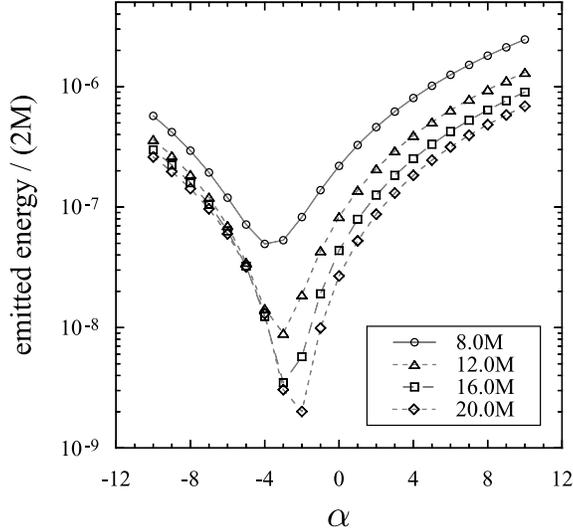}
\caption{%
The total energy emitted in gravitational waves ($l=2$) from the homogeneous dust collapse with magnetic field for four values of initial radius $r_{s0} = 8M$, $12M$, $16M$, and $20M$ as function of the magnetic field strength.
}%
\label{fig-alpha-E}
\end{figure}
\end{center}

In  Figure \ref{fig-alpha-E}, we show the total energy emitted in gravitational waves as a function of $\alpha$ for several values of the initial radius of the dust sphere. 
From this figure, we can see that the critical value $\alpha_c$,  increases as $r_{s0}$ increases, see also Table \ref{tab-energy}, while the emitted energy decreases as $r_{s0}$ increases.
However the effect of magnetic field  is stronger for the model with larger initial radius, i.e.,
the emitted energy varies with the strength of the magnetic field. Actually,
as we mentioned earlier the ratio of the  energy of the magnetized collapse over
the energy of the non-magnetized one varies from 0.22 ($\alpha=\alpha_c$) to 
11.2  ($\alpha=10$)  for $r_{s0}=8M$. For $r_{s0}=20M$ the same ratio varies
from 0.050 (at $\alpha=\alpha_c$) to 25.7 (at $\alpha=10$).

%
%

%
%
%
%
%
%
%
%
%
%
%

\subsection{Numerical results for the initial magnetic field profile (II)}
\label{sec:VI-2}

\begin{figure}[htbp]
\begin{center}
\includegraphics[height=7cm]{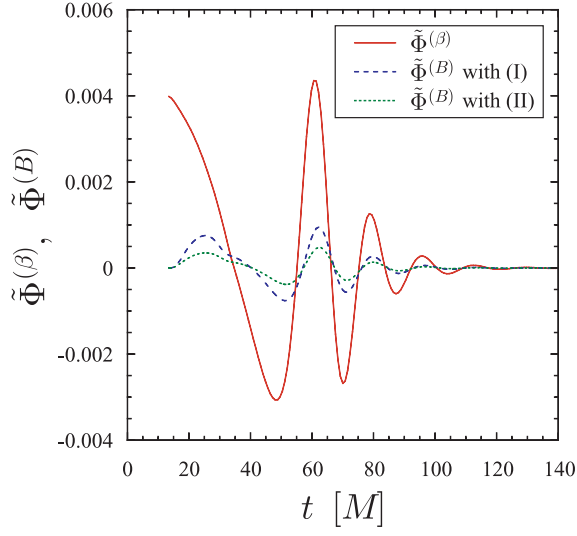}
\end{center}
\caption{
Waveforms of quadrupolar gravitational radiation from the magnetized OS collapse 
associated with $\tilde{\Phi}^{(\beta)}$ and $\tilde{\Phi}^{(B)}$, for the two 
initial profiles (I and II) for the magnetic field.
}
\label{fig:decompose1}
\end{figure}

The second magnetic field profile  considered in this work is the one that has its  maximum close to the stellar surface and is described by equation (\ref{MF2}).
Here, again we assumed an  initial profile for the fluid velocity that has $\bar{\beta}_0=const.$ and the initial radius of the dust spher is set to  $r_{s0}=8M$. 
In Fig. \ref{fig:decompose1}, we show the two components of the waveforms $\tilde{\Phi}^{(\beta)}$ and $\tilde{\Phi}^{(B)}$ for  profiles (I) and (II).
From this figure, we can see that the phase of $\tilde{\Phi}^{(B)}$ is almost independent of the initial profile of the magnetic field  while the  amplitude of the gravitational wave associated with the profile (II) is smaller than the one associated with the profile (I).
%
%
%

\begin{center}
\begin{figure}[htbp]
\includegraphics[height=7cm]{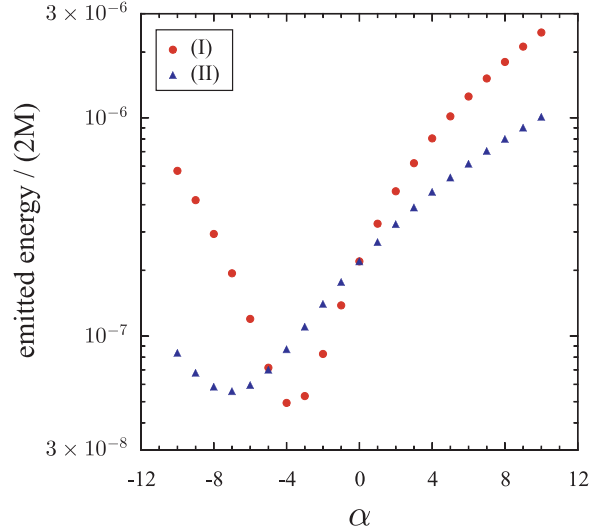}
\caption{
Total energy emitted in quadrupole gravitational waves from the homogeneous 
dust collapse with two different initial magnetic field profiles as functions
of the magnetic field strength.
The filled circles correspond to results for the profile (I)  
and the filled triangles for profile (II)}.
\label{fig-alpha-E-R08ii}
\end{figure}
\end{center}

In Fig. \ref{fig-alpha-E-R08ii}, we compare the total energy emitted in gravitational waves as functions of the magnetic filed strength for the two  initial magnetic field profiles. 
In this figure, we observe that the critical value $\alpha_c$, depends strongly on the initial profile of the magnetic field. 
Actually, for the profile (II)  we get $E_{GW,2}^{(B)}/(2M) = 3.267 \times 10^{-9}$, $C_{GW,2}/(2M) = 2.314 \times 10^{-8}$, and $\alpha_c = -7.08$.
These difference suggest that the form of the magnetic field affects in a critical way the amount of the emitted gravitational waves and worths more elaborate study.

%
%
%
%
%

\section{Conclusion}
\label{sec:VII}

In this paper, we have studied the influence of magnetic fields on the efficiency of collapse in emitting gravitational waves.
We have considered an interior Oppenheimer-Snyder solution describing collapsing dust and we studied how the amplitude and the waveform of the quadrupole axial perturbations were affected by the magnetic field, which actually enters as a second order perturbation term. These second order terms coming from the magnetic field are initially small but as the collapse proceeds they get amplified and become significant.  For this study we have assumed that the magnetic field is axially symmetric and the $l_M=1$ magnetic field perturbations are the ones that couple to the $l=2$ axial perturbations of the gravitational field. Additional assumptions have been made concerning the initial data and the influence of the magnetic field in the exterior.  That is, we assumed momentarily static initial data independent of the magnetic field and we have not taken into account the influence of the exterior magnetic field in the propagating gravitational waves. 

The main result of this study is the proof of the strong influence of the magnetic field in the gravitational wave luminosity during the collapse. Depending on the initial profile of the magnetic field and its strength the energy outcome can be easily up to one order higher than what we get from the non-magnetized collapse. Additionally, we observed that the initial profile of the magnetic field perturbations can affect the energy output while it is possible to observe important phase shifts induced by the presence of the magnetic field.  Since for a large initial radius the time  needed for the   black hole formation in longer, then the magnetic field acts for longer time on the collapsing fluid and its effect becomes more significant in the emitted gravitational wave signal.

Concluding, we believe that although this study might be considered as a "toy problem" it has most of the ingredients needed in emphasizing the importance of the magnetic fields in the study of the gravitational wave output during the collapse. It is obvious that 3D numerical MHD codes will provide the final answer to the questions raised here, but this work provides hints and raises issues that need to be studied.

\acknowledgments

We would like to thank Tomohiro Harada, Kenta Kiuchi, Hideki Maeda, and Kei-ichi Maeda
for helpful conversations. This work was supported in part by the Marie Curie Incoming International
Fellowships (MIF1-CT-2005-021979) and by the Grant-in-Aid for Scientific Research from the Ministry of
Education, Culture, Sports and Technology of Japan (Young Scientist (B) 17740155). S.Y. is supported by
the Grant-in-Aid for the 21st Century COE ``Holistic Research and Education Center for Physics of
Self-organization Systems'' from the Ministry of Education, Culture, Sports, Science and Technology of
Japan. KK acknowledges the support of the GSRT via the Pythagoras II program.
\appendix

\section{Perturbed energy-momentum tensor for the electromagnetic field inside the star}   
\label{sec:appendix_1}

The non-zero components of the perturbed energy-momentum tensor $\delta T_{\mu\nu}^{(EM)}$
for the electromagnetic field associated with dipole ($l_M=1$) perturbations inside the star are given by the following relations
\begin{align}
 \delta T_{\eta \eta}^{(EM)} &= \frac{1}{24\pi} \left[\frac{2({b_1}^2 + {b_3}^2)}{R^2 \sin^2\chi}
    + \frac{{b_2}^2}{R^2 \sin^4\chi}\right]P_0
    - \frac{1}{12\pi} \left[\frac{{b_1}^2 + {b_3}^2}{R^2 \sin^2\chi} - \frac{{b_2}^2}{R^2 \sin^4\chi}\right]P_2, \\
 \delta T_{\chi \chi}^{(EM)} &= \frac{1}{24\pi} \left[\frac{2({b_1}^2 + {b_3}^2)}{R^2 \sin^2\chi}
    - \frac{{b_2}^2}{R^2 \sin^4\chi}\right]P_0
    - \frac{1}{12\pi} \left[\frac{{b_1}^2 + {b_3}^2}{R^2 \sin^2\chi} + \frac{{b_2}^2}{R^2\sin^4\chi}\right]P_2, \\
 \delta T_{\chi \theta}^{(EM)} &= \frac{{b_1} {b_2}}{12\pi R^2 \sin^2\chi} (\partial_{\theta} P_2), \\
 \delta T_{\chi \phi}^{(EM)} &= -\frac{{b_2} {b_3}}{12\pi R^2 \sin^2\chi} \sin\theta (\partial_{\theta} P_2), \\
 \delta T_{\theta \theta}^{(EM)} &= \frac{{b_2}^2}{24\pi R^2 \sin^2\chi}P_0 \gamma_{\theta \theta}
    + \frac{{b_3}^2 - {b_1}^2}{12\pi R^2}Z_{\theta \theta}^{l=2}
    + \frac{{b_2}^2}{12\pi R^2 \sin^2\chi} P_2 \gamma_{\theta \theta}, \\
 \delta T_{\theta \phi}^{(EM)} &= \frac{{b_1}{b_3}}{12\pi R^2}(S_{\theta :\phi}+S_{\phi :\theta})^{(l=2)}, \\
 \delta T_{\phi \phi}^{(EM)} &= \frac{{b_2}^2}{24\pi R^2 \sin^2\chi}P_0 \gamma_{\phi \phi}
    + \frac{{b_3}^2 - {b_1}^2}{12\pi R^2}Z_{\phi \phi}^{l=2}
    + \frac{{b_2}^2}{12\pi R^2 \sin^2\chi} P_2 \gamma_{\phi \phi}.
\end{align}
Following \cite{Gundlach2000}, we can expand the perturbed energy-momentum tensor 
for polar parity perturbations in terms of tensor spherical harmonics as
\begin{align}
 \Delta t_{\mu\nu} \equiv   \left(
     \begin{array}{cc}
       \Delta t_{AB}P_{l} &   \Delta t_A^{\rm polar} P_{l:a}\\
       \Delta t_A^{\rm polar} P_{l:a} &  r^2 \Delta t^3 P_{l}\gamma_{ab} + \Delta t^2 Z_{ab}^{l}
     \end{array}\right)\,,
\end{align}
where $Z^{l}_{ab}\equiv P_{l:ab}+l(l+1)P_{l}\gamma_{ab}/2$. 
[See, for the axial parity perturbations, equation (\ref{AFP}).] 
Then, it has been found that the non-zero tensor-harmonic expansion coefficients of $\delta T_{\mu\nu}^{(EM)}$ for  $l_M=1$ are coupled with the $l=0$ and $l=2$ perturbations. 

The expansion coefficients for the $l=0$ perturbations are the following:
\begin{align}
 \Delta t_{\eta \eta} &= \frac{1}{24\pi}\left[\frac{2({b_1}^2 + {b_3}^2)}{R^2 \sin^2\chi}
     + \frac{{b_2}^2}{R^2 \sin^4\chi}\right],\\
 \Delta t_{\chi \chi} &= \frac{1}{24\pi}\left[\frac{2({b_1}^2 + {b_3}^2)}{R^2 \sin^2\chi}
     - \frac{{b_2}^2}{R^2 \sin^4\chi}\right],\\
 \Delta t^3    &= \frac{{b_2}^2}{24\pi R^4 \sin^4\chi}\, ,
\end{align}
while the coefficients for the $l=2$ perturbations are:
\begin{gather}
 \Delta t_{\chi}^{(a)} = -\frac{{b_2}{b_3}}{12\pi R^2 \sin^2\chi},\ \
 \Delta t         = \frac{{b_1}{b_3}}{12\pi R^2}, \label{AA}\\
 \Delta t_{\eta \eta}  = \frac{1}{12\pi} \left[\frac{{b_2}^2}{R^2 \sin^4\chi} - \frac{{b_1}^2
     + {b_3}^2}{R^2 \sin^2\chi}\right],\ \
 \Delta t_{\chi \chi}  = - \frac{1}{12\pi} \left[\frac{{b_2}^2}{R^2 \sin^4\chi} + \frac{{b_1}^2
     + {b_3}^2}{R^2 \sin^2\chi}\right], \label{PP1} \\
 \Delta t_{\chi}^{(p)} = \frac{{b_1}{b_2}}{12\pi R^2 \sin^2\chi},\ \
 \Delta t^2       = \frac{{b_3}^2 - {b_1}^2}{12\pi R^2},\ \
 \Delta t^3       = \frac{{b_2}^2}{12\pi R^4 \sin^4\chi}, \label{PP2}
\end{gather}
where (\ref{AA}) belongs to axial parity perturbations, 
and (\ref{PP1}) and (\ref{PP2}) belong to polar parity perturbations.

\section{Perturbed energy-momentum tensor for the electromagnetic field outside the star}   
\label{sec:appendix_2}

The non-zero components of the perturbed energy-momentum tensor $\delta T_{\mu\nu}^{(EM)}$
for the electromagnetic field associated with dipole ($l_M=1$) perturbations outside the star are given by the following relations
\begin{align}
 \delta \tilde{T}_{tt}^{(EM)} =&
    \frac{1}{24\pi} \left[f \left(\tilde{e}_2^{\ 2} + \frac{\tilde{b}_2^{\ 2}}{r^4}\right)
    + \frac{2}{r^2} \left\{\tilde{e}_1^{\ 2} + \tilde{e}_3^{\ 2} + f^2 (\tilde{b}_1^{\ 2} + \tilde{b}_3^{\ 2})\right\}
    \right]P_0 \nonumber \\
    &
    + \frac{1}{12\pi} \left[f \left(\tilde{e}_2^{\ 2} + \frac{\tilde{b}_2^{\ 2}}{r^4}\right)
    - \frac{1}{r^2} \left\{\tilde{e}_1^{\ 2} + \tilde{e}_3^{\ 2} + f^2 (\tilde{b}_1^{\ 2} + \tilde{b}_3^{\ 2})\right\}
    \right]P_2, \\
 \delta \tilde{T}_{tr}^{(EM)} =&
    \frac{\tilde{e}_1 \tilde{b}_1 + \tilde{e}_3 \tilde{b}_3}{6\pi r^2} P_0
    - \frac{\tilde{e}_1 \tilde{b}_1 + \tilde{e}_3 \tilde{b}_3}{6\pi r^2} P_2, \\
 \delta \tilde{T}_{t\theta}^{(EM)} =&
    \frac{1}{12\pi} \left(-f \tilde{e}_2 \tilde{b}_3 + \frac{1}{r^2} \tilde{e}_1 \tilde{b}_2 \right)
    (\partial_{\theta} P_2), \\
 \delta \tilde{T}_{t\phi}^{(EM)} =&
    -\frac{1}{12\pi} \left(f \tilde{e}_2 \tilde{b}_1 + \frac{1}{r^2} \tilde{e}_3 \tilde{b}_2 \right)
    \sin\theta (\partial_{\theta} P_2), \\
 \delta \tilde{T}_{rr}^{(EM)} =&
    \frac{1}{24\pi} \left[-\frac{1}{f} \left(\tilde{e}_2^{\ 2} + \frac{\tilde{b}_2^{\ 2}}{r^4}\right)
    + \frac{2}{f^2r^2} \left\{\tilde{e}_1^{\ 2} + \tilde{e}_3^{\ 2} + f^2 (\tilde{b}_1^{\ 2} + \tilde{b}_3^{\ 2})\right\}
    \right]P_0 \nonumber \\
    &
    + \frac{1}{12\pi} \left[-\frac{1}{f} \left(\tilde{e}_2^{\ 2} + \frac{\tilde{b}_2^{\ 2}}{r^4}\right)
    - \frac{1}{f^2r^2} \left\{\tilde{e}_1^{\ 2} + \tilde{e}_3^{\ 2} + f^2 (\tilde{b}_1^{\ 2} + \tilde{b}_3^{\ 2})\right\}
    \right]P_2, \\
 \delta \tilde{T}_{r\theta}^{(EM)} =&
    \frac{1}{12\pi} \left(-\frac{1}{f} \tilde{e}_2 \tilde{e}_3 + \frac{1}{r^2} \tilde{b}_1 \tilde{b}_2 \right)
    (\partial_{\theta} P_2), \\
 \delta \tilde{T}_{r\phi}^{(EM)} =&
    -\frac{1}{12\pi} \left(\frac{1}{f} \tilde{e}_1 \tilde{e}_2 + \frac{1}{r^2} \tilde{b}_2 \tilde{b}_3 \right)
    \sin\theta (\partial_{\theta} P_2), \\
 \delta \tilde{T}_{\theta \theta}^{(EM)} =&
    \frac{r^2}{24\pi} \left(\tilde{e}_2^{\ 2} + \frac{\tilde{b}_2^{\ 2}}{r^4}\right) P_0 \gamma_{\theta \theta}
    + \frac{1}{12\pi f} \left[\tilde{e}_1^{\ 2} - \tilde{e}_3^{\ 2}
    - f^2 (\tilde{b}_1^2 - \tilde{b}_3^{\ 2})\right]Z_{\theta \theta}^{(l=2)}
    + \frac{r^2}{12\pi} \left(\tilde{e}_2^{\ 2} + \frac{\tilde{b}_2^{\ 2}}{r^4}\right) P_2 \gamma_{\theta \theta}, \\
 \delta \tilde{T}_{\theta \phi}^{(EM)} =&
    \frac{1}{12\pi} \left(-\frac{1}{f} \tilde{e}_1 \tilde{e}_3 + f \tilde{b}_1 \tilde{b}_3 \right)
    (S_{\theta :\phi} + S_{\phi :\theta})^{(l=2)}, \\
 \delta \tilde{T}_{\phi \phi}^{(EM)} =&
    \frac{r^2}{24\pi} \left(\tilde{e}_2^{\ 2} + \frac{\tilde{b}_2^{\ 2}}{r^4}\right) P_0 \gamma_{\phi \phi}
    + \frac{1}{12\pi f} \left[\tilde{e}_1^{\ 2} - \tilde{e}_3^{\ 2}
    - f^2 (\tilde{b}_1^2 - \tilde{b}_3^{\ 2})\right]Z_{\phi \phi}^{(l=2)}
    + \frac{r^2}{12\pi} \left(\tilde{e}_2^{\ 2} + \frac{\tilde{b}_2^{\ 2}}{r^4}\right) P_2 \gamma_{\phi \phi}.
\end{align}
As for the interior, the non-zero expansion coefficients of $\delta T_{\mu\nu}^{(EM)}$ for  $l_M=1$ electromagnetic field perturbations are coupled with the $l=0$ and $l=2$ perturbations. 

The expansion coefficients for the $l=0$ perturbations are: 
\begin{align}
 \Delta \tilde{t}_{tt} =&
    \frac{1}{24\pi} \left[f \left(\tilde{e}_2^{\ 2} + \frac{\tilde{b}_2^{\ 2}}{r^4}\right)
    + \frac{2}{r^2} \left\{\tilde{e}_1^{\ 2} + \tilde{e}_3^{\ 2} + f^2 (\tilde{b}_1^2 + \tilde{b}_3^{\ 2})\right\}
    \right], \\
 \Delta \tilde{t}_{tr} =&
    \frac{\tilde{e}_1 \tilde{b}_1 + \tilde{e}_3 \tilde{b}_3}{6\pi r^2}, \\
 \Delta \tilde{t}_{rr} =&
    \frac{1}{24\pi} \left[-\frac{1}{f}\left(\tilde{e}_2^{\ 2} + \frac{\tilde{b}_2^{\ 2}}{r^4}\right)
    + \frac{2}{f^2 r^2} \left\{\tilde{e}_1^{\ 2} + \tilde{e}_3^{\ 2} + f^2 (\tilde{b}_1^2 + \tilde{b}_3^{\ 2})\right\}
    \right], \\
 \Delta \tilde{t}^{3} =&
    \frac{r^2}{24\pi} \left(\tilde{e}_2^{\ 2} + \frac{\tilde{b}_2^{\ 2}}{r^4}\right)\, ,
\end{align}
while the coefficients for the $l=2$ polar parity perturbations have the form: 
\begin{align}
 \Delta \tilde{t}_{tt} =&
    \frac{1}{12\pi} \left[f \left(\tilde{e}_2^{\ 2} + \frac{\tilde{b}_2^{\ 2}}{r^4}\right)
    - \frac{1}{r^2} \left\{\tilde{e}_1^{\ 2} + \tilde{e}_3^{\ 2} + f^2 \left(\tilde{b}_1^{ 2} + \tilde{b}_3^{\ 2}\right)
    \right\}\right], \\
 \Delta \tilde{t}_{tr} =&
    -\frac{\tilde{e}_1 \tilde{b}_1 + \tilde{e}_3 \tilde{b}_3}{6 \pi r^2}, \\
 \Delta \tilde{t}_{rr} =&
    \frac{1}{12\pi} \left[-\frac{1}{f} \left(\tilde{e}_2^{\ 2} + \frac{\tilde{b}_2^{\ 2}}{r^4}\right)
    - \frac{1}{f^2 r^2} \left\{\tilde{e}_1^{\ 2} + \tilde{e}_3^{\ 2} + f^2 \left(\tilde{b}_1^{ 2} + \tilde{b}_3^{\ 2}\right)
    \right\}\right], \\
 \Delta \tilde{t}_{t}^{(p)} =&
    \frac{1}{12\pi} \left(-f \tilde{e}_2 \tilde{b}_3 + \frac{1}{r^2} \tilde{e}_1 \tilde{b}_2\right), \\
 \Delta \tilde{t}_{r}^{(p)} =&
    \frac{1}{12\pi} \left(-\frac{1}{f} \tilde{e}_2 \tilde{e}_3 + \frac{1}{r^2} \tilde{b}_1 \tilde{b}_2\right), \\
 \Delta \tilde{t}^{2} =&
    \frac{1}{12\pi f} \left[\tilde{e}_1^{\ 2} - \tilde{e}_3^{\ 2} - f^2 (\tilde{b}_1^{\ 2} - \tilde{b}_3^{\ 2})\right], \\
 \Delta \tilde{t}^{3} =&
    \frac{1}{12\pi} \left(\tilde{e}_2^{\ 2} + \frac{\tilde{b}_2^{\ 2}}{r^4}\right)\, .
\end{align}
Finally, the coefficients for the $l=2$ axial parity perturbations have the form:
\begin{align}
 \Delta \tilde{t}_t^{(a)} =&
    -\frac{1}{12\pi} \left(f \tilde{e}_2 \tilde{b}_1 + \frac{1}{r^2} \tilde{e}_3 \tilde{b}_2\right), \\
 \Delta \tilde{t}_r^{(a)} =&
    -\frac{1}{12\pi} \left(\frac{1}{f} \tilde{e}_1 \tilde{e}_2 + \frac{1}{r^2} \tilde{b}_2 \tilde{b}_3\right), \\
 \Delta \tilde{t} =&
    -\frac{1}{12\pi} \left(\frac{1}{f} \tilde{e}_1 \tilde{e}_3 - f \tilde{b}_1 \tilde{b}_3 \right)\,.
\end{align}


\end{document}